\documentclass[aastex]{article}
\usepackage[onecolumn]{priya}

\lefthead{Crittenden, Natarajan, Pen \& Theuns}
\righthead{Spin Induced Galaxy Alignments}

\def\gs{\mathrel{\raise0.35ex\hbox{$\scriptstyle >$}\kern-0.6em 
les
\lower0.40ex\hbox{{$\scriptstyle \sim$}}}}
\def\ls{\mathrel{\raise0.35ex\hbox{$\scriptstyle <$}\kern-0.6em 
ggles
\lower0.40ex\hbox{{$\scriptstyle \sim$}}}}
\def\ltorder{
\mathrel{\raise.3ex\hbox{$<$}\mkern-14mu\lower0.6ex\hbox{$\sim$}}
}
\def\gtorder{
\mathrel{\raise.3ex\hbox{$>$}\mkern-14mu\lower0.6ex\hbox{$\sim$}}
}

\def\ep{\epsilon}
\def\epp{\epsilon_{+}}
\def\epc{\epsilon_{\times}}

\input psfig
\begin{document}
\title{Spin Induced Galaxy Alignments and
their implications for weak lensing measurements}
\author{Robert G. Crittenden$^1$, Priyamvada Natarajan$^{2,3}$, 
Ue-Li Pen$^4$ 
\& Tom Theuns$^2$}
\affil{1 Department of Applied Mathematics and Theoretical Physics, 
Wilberforce Road, Cambridge CB3 0WA, UK}
\affil{2 Institute of Astronomy, Madingley Road, Cambridge CB3 0HE, UK}
\affil{3 Department of Astronomy, Yale University, New Haven, CT, USA}
\affil{4 CITA, McLennan Labs, University of Toronto, Toronto, M5S 3H8}

\begin{abstract}

Large scale correlations in the orientations of galaxies can result
from alignments in their angular momentum vectors.  These alignments
arise from the tidal torques exerted on neighboring proto-galaxies by
the smoothly varying shear field.  We compute the predicted amplitude
of such ellipticity correlations using the Zel'dovich approximation
for a realistic distribution of galaxy shapes.  Weak gravitational
lensing can also induce ellipticity correlations since the images of
neighboring galaxies will be distorted coherently.  On comparing
these two effects that induce shape correlations, we find that for
current weak lensing surveys with a median redshift of $z_m=1$, the
intrinsic signal is of order 1-10\% of the measured signal.
However, for shallower surveys with $z_m \le 0.3$, the intrinsic
correlations dominate over the lensing signal.  The distortions
induced by lensing are curl-free, whereas those resulting from
intrinsic alignments are not.  This difference can be used to
disentangle these two sources of ellipticity correlations.

\end{abstract}

\section{Introduction}

Gravitational lensing can be used to map the detailed distribution of
matter in the Universe over a range of scales (Gunn 1967). Systematic
distortions in the shapes and orientations of high redshift background
galaxies are induced by mass inhomogeneities along the line of
sight. In strong lensing, a single massive foreground cluster will
cause background galaxies to be significantly magnified and distorted.
Weak lensing, on the other hand, measures the cumulative effect of
less massive systems along the line of sight statistically (Gunn 1967;
Blandford et al. 1991; Miralda-Escude 1991; Kaiser 1992; see a recent
review by Bartelmann \& Schneider 1999).

The lensing effect depends only on the projected surface mass density
and is independent of the luminosity or the dynamical state of the
mass distribution. Thus, this technique can potentially provide
invaluable constraints on the distribution of matter in the Universe
and the underlying cosmological model (Bernardeau, van Waerbeke \&
Mellier 1997; van Waerbeke, Bernardeau \& Mellier 1999).  There has
been considerable progress in theoretical calculations of the effects
of weak lensing by large-scale structure, both analytically and using
ray-tracing through cosmological N-body simulations (Kaiser 1992;
Bernardeau, van Waerbeke \& Mellier 1997; Jain \& Seljak 1997; Jain,
Seljak \& White 2000).

Recently, several teams have reported observational detections of
`cosmic shear' -- weak lensing on scales ranging from an arc-minute to
ten arc-minutes (Van Waerbeke et al. 2000; Bacon, Refregier \& Ellis
2000; Wittman et al. 2000; Kaiser, Wilson \& Luppino 2000). At
present, these studies are limited by observational effects, such as
shot noise due to the finite number of galaxies and the accuracy with
which shapes can actually be measured given the optics and seeing
(Kaiser 1995; Bartelmann \& Schneider 1999; Kuijken 1999).  In
addition, the intrinsic ellipticity distribution of galaxies and their
redshift distribution is still somewhat uncertain.  These
observational difficulties can be potentially overcome with more data.

However, an important theoretical issue remains. In modeling the
distortion produced by lensing, it is assumed that the a priori
intrinsic correlations in the shapes and orientations of background
galaxies are negligible. Correlations in the intrinsic ellipticities
of neighboring galaxies are expected to arise from the galaxy
formation process, for example as a consequence of correlations
between the angular momenta of galaxies when they assemble.  We
compute the strength of these correlations in linear theory, in the
context of Gaussian initial fluctuations.

To do so, we approximate the projected shape of a galaxy on the sky by
an ellipsoid with semi-axes $a$, $b$ ($a > b$).  The orientation of
the ellipsoid depends on the angle $\psi$ between the major axis and
the chosen coordinate system, while its magnitude is given by $|\ep| =
(a^2 - b^2)/(a^2 + b^2)$. Both the magnitude of the ellipticity and
its orientation can be concisely described by the complex quantity
$\epsilon^{(o)}$,
\begin{eqnarray} 
\epsilon^{(o)} = |\epsilon^{(o)}| e^{2i\psi} 
= [\epsilon_{+}^{(o)} + i \epsilon_{\times}^{(o)}].
\end{eqnarray} 
where the superscript $^{(o)}$ denotes the 
observed shape. 

In the linear regime and under the assumption of weak lensing, the
lensing equation can be written as,
\begin{eqnarray}
\epsilon^{(o)}\,=\,\frac{\epsilon\,+\,g}{1\,+\,g^{*}\epsilon},
\end{eqnarray}
where $g$ is the complex shear and $\epsilon$ the intrinsic shape of
the source (Kochanek 1990; Miralda-Escude 1991). Furthermore, in the weak regime, correlations of this
distortion field are
\begin{eqnarray}
\langle{\epsilon^{(o)}}(\mathbf{x_1})\,{\epsilon^{(o)*}}(\mathbf{x_2})
\rangle\,\simeq\,
\langle{\epsilon (\mathbf{x_1})}\,{\epsilon^{*}}(\mathbf{x_2})\rangle\,+
\,\langle{g^{*}}(\mathbf{x_2}){\epsilon
(\mathbf{x_1})}+{g(\mathbf{x_1})}
{\epsilon^{*}}(\mathbf{x_2})\rangle
\,+\,\langle{g(\mathbf{x_1})}{g^{*}}(\mathbf{x_2})\rangle\,
\end{eqnarray}
where the $^*$ denotes complex conjugation.\footnote{ 
In the remainder of the text 
we will write two-point correlations functions in the following form:
$\langle f\,g{'} \rangle\,\equiv
\langle f(\mathbf{x_1})\,g(\mathbf{x_2})\rangle\,$.}
In this paper we will examine the first term,
which arises from intrinsic shape correlations.  Previous analyses
have focused on the third term of this expression, correlations due to
weak lensing.  The second term, which is due to correlations between
the lensing galaxies and the intrinsic shapes of the galaxies being
lensed, will not be addressed here. Naively however, we expect
this contribution to be small, since  
the mean distance between the lensing  and lensed galaxies far  
exceeds the distance scale over which angular momentum correlations are 
important.

We will assume that shape correlations arise primarily from
correlations in the direction of the angular momentum vectors of
neighboring galaxies.  Spiral galaxies are disk-like with the angular
momentum vector perpendicular to the plane of the disk, so that
angular momentum couplings will be translated into shape correlations.
We will assume that for ellipticals the angular momentum vector also
lies along its shortest axis on average, as it does for the spirals.
However, since elliptical galaxies are intrinsically more round, the
correlation amplitude will be smaller.  Below we will use the observed
ellipticity distributions of each morphological type in the
computation of the shape correlations.  For weak lensing, in contrast,
the induced shape correlations are independent of the original shapes
of the lensed galaxies.  In the next sub-sections we will briefly
review the origin of angular momentum and recent work on understanding
intrinsic ellipticity correlations.

\subsection{Origins of Angular Momentum}

The angular momentum of the matter contained in a volume $V$ is
defined as,
\begin{eqnarray}
{\mathbf {L}}(t) = \int_{V}[{\mathbf {r}}(t) - {\bar
{\mathbf r}}(t)] \times {\mathbf {v}}({\mathbf {r}}, t)\,\rho({\mathbf{r}},t)
\,d^3 r,
\end{eqnarray}
where ${\bar {\mathbf {r}}}$ is the center of mass and
$\rho({\mathbf{r}},t)$ is the density.  Hoyle (1949) suggested that
the origin of galactic angular momentum is tidal torquing between the
proto-galaxy and the surrounding matter distribution.  Most of the
angular momentum of an object is imparted before the over-dense region
completely collapses. After collapse, tidal torquing will be
inefficient and the object will simply conserve its spin.

Peebles (1969) used perturbation theory to calculate the growth rate
of angular momentum contained within a comoving spherical region. For
such a spherical region, there are no torques initially, so the growth
occurs at second order as a result of convective effects on the
bounding surface.  In contrast, Doroshkevich (1970) showed that the
angular momentum of a proto-galaxy grows at first order since, in
general, proto-galactic regions are not spherical, generating an
initial tidal torque.  White (1984) described this process using the
Zel'dovich (1970) approximation and showed that the spin grows
linearly in time, for an Einstein-de Sitter universe.

Following White (1984), we consider the growth of fluctuations in an
expanding Friedmann Universe filled with pressure-free dust ($p = 0$)
in Lagrangian perturbation theory.  The trajectory of a dust particle
can be written in comoving coordinates ${\mathbf{x}}={\mathbf{r}}/a$
in terms of the gradient of the gravitational potential $\Psi$,
${\mathbf{x}}({\mathbf {q}},t) = {\mathbf{q}} - D(t) \nabla \Psi$
(Zeldovich 1970).  $D(t)$ describes the growth of modes in linear
theory and is proportional to the cosmological expansion factor,
$a(t)$, for an Einstein de-Sitter model.  In terms of the Lagrangian
coordinates ${\mathbf{q}}$, the expression for angular momentum
becomes
\begin{eqnarray}
{\mathbf {L}}(t) = \rho_0 a^5 \int_{V_L}[{\mathbf {x}} - 
{\bar{\mathbf {x}}}] \times 
{\mathbf {\dot x}}\,\,d^3 q \simeq \rho_0 a^5 \int_{V_L}[{\mathbf {q}} - 
{\bar{\mathbf {q}}}] \times 
{\mathbf {\dot x}}\,\,d^3 q,
\end{eqnarray}
where $\rho_0$ is the present mean matter density and $V_L$ is the 
Lagrangian volume that corresponds to $V$. 
The latter expression is correct to second order since
${\mathbf {\dot x}} = - \dot {D}(t) \nabla \Psi$ is parallel 
to the displacement. 

We can progress by expanding the gradient of the gravitational
potential in a Taylor series around the center of mass,
\begin{eqnarray}
\partial_{\alpha}\Psi({\mathbf {q}}) \simeq
 \partial_{\alpha}\Psi({\bar{\mathbf {q}}}) + 
({\mathbf {q}} - {\bar{\mathbf {q}}})_\beta
T_{\alpha \beta} 
\end{eqnarray}
where the shear tensor is defined as the
second derivative of the gravitational potential,
$T_{\alpha \beta} ({\mathbf
q})\,=\,\partial_{\alpha}\,\partial_{\beta}\,\Psi({\mathbf q}).$
The angular momentum of a collapsing proto-galactic region before
turnaround then is given by,
\begin{eqnarray}
L_{\alpha}\,=\,a^2(t)\,{\dot D}(t)\,\epsilon_{\alpha \beta \gamma}\,
T_{\beta \sigma}\,I_{\sigma \gamma},
\end{eqnarray}
$I_{\sigma \gamma}$ is the moment of inertia of the
matter in the collapsing volume,  
\begin{eqnarray}
I_{\sigma \gamma} = \rho_0 a^3(t)\int_{V_L} ({\mathbf {q}} - 
{\bar{\mathbf {q}}})_\sigma ({\mathbf {q}} - {\bar{\mathbf {q}}})_\gamma d^3q.
\end{eqnarray}
Note that the volume element that initially contains the matter is in
fact much larger than that of the final galaxy in comoving
coordinates. In this picture, the angular momentum is constant after
turnaround.

This formalism has been used to study how angular momentum arises during
galaxy formation.  Heavens and Peacock (1988) used Eulerian
perturbation theory to compute the modulus of the angular momentum for
galaxies, assuming the object to form at the peak of a Gaussian
field (Bardeen et al. 1986).  They found that there is a broad
distribution in the angular momenta of collapsed objects, which is
only weakly correlated with the heights of the density peaks around
which galaxies form.

Catelan \& Theuns (1996a [CT96a]) expanded on this, working in
Lagrangian space instead of Eulerian space.  The results from these
two approaches are very similar, but the resulting expressions are
simpler for the Lagrangian case.  CT96a approximated the shape of the
object in Lagrangian space by an ellipsoid, which allowed the study of
how angular momentum was correlated with other aspects of the matter
distribution, such as its mass or its prolateness.  The results of
this analysis allows one to compute joint probability distributions,
for example between the mass and spin of a halo. These were found to
be in good agreement with the results from numerical simulations
(Sugerman, Summers and Kamionkowski 2000).

Extending their approach, Catelan and Theuns (1996b) used second-order
Lagrangian perturbation theory to estimate the contribution of
non-linear effects, which they showed to be small. They also
investigated the consequences of non-Gaussian primordial perturbations
(Catelan \& Theuns 1997), which they showed could have a significant
effect on galactic spins.

Lee \& Pen (2000) re-examined the origin of angular momentum on galaxy
scales and studied the statistics of both the magnitude and the
direction of the present day spin distribution using numerical
simulations.  They developed a method to reconstruct the gravitational
field, using only the direction of the angular momenta, since the
predictions for its magnitude have a large variance.  A central issue
in determining the magnitude of $L$ is the degree of correlation
between the principal axes of the inertia tensor and gravitational
shear tensor.  CT96a attempted to take this into account around peaks,
and found that such a correlation reduces the angular momentum by a
small factor.  Lee \& Pen (2000) demonstrated using numerical
simulations that this factor is in fact non-negligible.  They however
conclude that the approximation made by CT96a is adequate for
determining the direction of the angular momentum vector but not for
the magnitude.  Here we extend the treatment of Lee \& Pen (2000) and
much of our notation and formalism follows their paper.

\subsection{Intrinsic Ellipticity Correlations} 

There have been a number of preprints on this subject recently.  We
briefly review some of the results obtained by other groups here and
we will compare our calculations and results with these in more detail
in subsequent sections.

Two groups, Heavens et al. (2000) and Croft \& Metzler (2000), have
attempted to measure the strength of intrinsic correlations from high
resolution cosmological N-body simulations (that evolve only the dark
matter component) of the Virgo collaboration (Jenkins et al. 1998;
Thomas et al. 1998; Pearce et al. 1999). Some assumption must be made
to relate the dark matter halos in numerical simulations to the
expected ellipticity of the luminous galaxies that form within them.
Croft \& Metzler (2000) measure the projected ellipticities of dark
matter halos and the correlation of pairs as a function of
separation. They then assumed that halo shapes are synonymous with
galaxy shapes, and having done so claim to find a positive signal for
the correlation on scales of the order of 20 h$^{-1}$ Mpc (limited by
the largest box size available). The results obtained in three
dimensions were then projected into two dimensional angular
ellipticity correlation functions, taking into account the viewing
angle.  They compute the induced correlations in the ellipticity and
compare to recent reported measurements of the observed lensing
signal.  While there is a large uncertainty arising due to the unknown
redshift distribution of the sheared background galaxies, they find
that at most 10 - 20\% of the measured signal could be attributed to
contamination from residual intrinsic correlations.

Heavens et al. (2000) have studied correlations in the intrinsic
shapes of spiral galaxies also using the Virgo simulations.  However,
they use the angular momentum of the halo (rather than the actual
shape as done by Croft \& Metzler) and assume that its direction is
perpendicular to that of a thin disk.  They compute the 3-D
ellipticity correlation function and its 2-D projection directly from
simulations populated by $\sim 10^5$ halos. They also conclude on
comparing with recent measurements of the shear induced by lensing on
large scales that the contamination from intrinsic correlations is
small on most angular scales of interest - the contamination is
roughly at the 10-20\% level on scales of $0.1 - 10'$.

Catelan, Kamionkowski \& Blandford (2000) have recently presented an
analytic calculation to assess the importance of intrinsic galaxy
shape alignments and the consequent mimicking of the signal produced
by weak gravitational lensing.  They make the {\it Ansatz} that the
ellipticity is linearly proportional to the tidal shear and 
calculate correlations due to intrinsic shape correlations as a
function of scale.  (While originally meant to apply to ellipticity 
correlations resulting from angular momentum couplings, Catelan et al. 
now use this {\it Ansatz} only for ellipticities induced by the halo shapes.
[M. Kamionkowski, private communication.])  They also consider possible 
means of discriminating the lensing signal from intrinsic allignments.  
 
Very recently, the first observational detection of the  
magnitude of spin-spin correlations 
has been reported by Pen, Lee \& Seljak
(2000).  They construct the simplest quadratic two-point spin-spin
correlation function in the context of linear perturbation theory and
compare the statistic computed for galaxies in the Tully catalog. They
claim a detection at the 97\% confidence level out to a few Mpc.

Several authors have pointed out that one of the important
discriminants between the correlations arising due to lensing versus
those from intrinsic alignments is the prediction of the existence of
non-zero `B-type' curl modes in the shear field in the intrinsic case
(Kaiser 1992; Stebbins 1996; Kamionkowski et al. 1998).  A detailed
decomposition of the shear field into the `B' and `E', or pure
gradient, modes for intrinsic correlations is presented in Crittenden
et al. (2000).

\subsection{Schematic Outline}

Our goal is to calculate the two point correlation of the intrinsic
shape distribution of galaxies, $\langle \epsilon \epsilon{'}
\rangle$, as a function of projected distance.  Since the following
calculation is quite complex, we present a brief schematic outline to
guide the reader and to clarify the simplifying assumptions that we
make.  Our approach is primarily analytic, but we also use numerical
realizations of Gaussian fields to verify some of our results.

The intrinsic ellipticity of a galaxy depends on its three dimensional
shape, its orientation and on the direction of its angular momentum,
$\ep \equiv \ep({\mathbf{S}}, {\mathbf{\hat{L}}})$.  Here we use
${\mathbf{S}}$ to denote the shape and orientation degrees of
freedom. We will implicitly assume that the galaxy is ellipsoidal and
that its angular momentum lies parallel to the shortest axis of the
ellipsoid.  The expected correlation between ellipticities at
different points is
\begin{eqnarray} 
\langle \ep \ep{'^*} \rangle = \int d{\mathbf{S}}\, d{\mathbf{S}}{'}\,  
d{\mathbf{\hat{L}}}\, d{\mathbf{\hat{L}}}{'}\,  \ep({\mathbf{S}}, 
{\mathbf{\hat{L}}}) \ep({\mathbf{S}}{'}, {\mathbf{\hat{L}}}{'}) 
{\cal{P}}({\mathbf{S}}, {\mathbf{\hat{L}}}, {\mathbf{S}}{'}, 
{\mathbf{\hat{L}}}{'}),
\end{eqnarray}
where ${\cal P}$ denotes the joint probability distribution.  
The present three-dimensional shapes of galaxies, quantified via their
axis ratios, are primarily determined by `local' processes like the extent of
dissipation within the collapsing dark matter halo. 
Thus, we will assume that
they are uncorrelated between neighboring galaxies, so that we can 
rewrite, ${\cal{P}}({\mathbf{S}}, {\mathbf{\hat{L}}},
{\mathbf{S}}{'}, {\mathbf{\hat{L}}}{'}) =
{\cal{P}}({\mathbf{\hat{L}}}, {\mathbf{\hat{L}}}{'}) {\cal{P}} (
{\mathbf{S}}) {\cal{P}}({\mathbf{S}}{'})$.  For each galaxy, we can
then integrate over all possible shapes and orientations to find the
average ellipticity of a galaxy with angular momentum in a given direction, 
$\bar{\ep}({\mathbf{\hat{L}}}) = \int
d{\mathbf{S}}\, \ep({\mathbf{S}},{\mathbf{\hat{L}}})
{\cal{P}}({\mathbf{S}})$.  This integration is described in detail
in Section 2.  

The resulting correlation is then simply given by,
\begin{eqnarray} 
\langle \ep \ep{'^*} \rangle = \int 
d{\mathbf{\hat{L}}}\, d{\mathbf{\hat{L}}}{'} \, \bar{\ep}(
{\mathbf{\hat{L}}}) \bar{\ep}({\mathbf{\hat{L}}}{'}) 
{\cal{P}}({\mathbf{\hat{L}}}, {\mathbf{\hat{L}}}{'}). 
\end{eqnarray}
To proceed we need to understand the correlations between the
directions of the angular momentum vectors of galaxies, or
explicitly the nature of ${\cal{P}}({\mathbf{\hat{L}}},
{\mathbf{\hat{L}}}{'}).$  Rather than attempt to calculate the 
angular momentum correlations directly, we instead relate them to 
correlations in the shear tensor, ${\mathbf{T}}$, 
which yields itself more easily to linear theory. 

As discussed above, the angular momentum of a given galaxy depends on
the tidal field and the moment of inertia, ${\mathbf{I}}$, of all the
matter that has turned around and which will eventually collapse to
form the galaxy.  To compute ${\cal{P}}({\mathbf{\hat{L}}},
{\mathbf{\hat{L}}}{'})$, one needs the full joint probability
function, ${\cal{P}}({\mathbf{I}},{\mathbf{T}},{\mathbf{I}}{'},
{\mathbf{T}}{'})$.  We make the simplifying assumption that the moment
of inertia at a given point is significantly correlated only with the
shear at that point, so that
${\cal{P}}({\mathbf{I}},{\mathbf{T}},{\mathbf{I}}{'} ,{\mathbf{T}}{'})
= {\cal{P}}({\mathbf{I}}|{\mathbf{T}})
{\cal{P}}({\mathbf{I}}{'}|{\mathbf{T}}{'})
{\cal{P}}({\mathbf{T}},{\mathbf{T}}{'})$.  Given a form for
${\cal{P}}({\mathbf{I}}|{\mathbf{T}})$, one can derive
${\cal{P}}(\hat{\mathbf{L}}|{\mathbf{T}})$.  However, since the local
stress tensor depends on the details of the mass distribution outside
the collapsed object as well, ${\cal{P}}({\mathbf{I}}|{\mathbf{T}})$
is not accurately known.  In Section 3, we follow LP00 and assume that
${\cal{P}}(\hat{\mathbf{L}}|{\mathbf{T}})$ is Gaussian, and use the
most general form that the correlation matrix could have as a function
of the shear tensor.  This allows us to derive an expression for
$\bar{\ep}({\mathbf{T}}) = \int d{\mathbf{\hat{L}}}
\ep({\mathbf{\hat{L}}}) {\cal{P}}({\mathbf{\hat{L}}}| {\mathbf{T}})$,
the expected mean ellipticity for a given shear tensor.

With these assumptions, the ellipticity correlation depends only on
how the tidal field is correlated from place to place,
\begin{eqnarray} 
\langle \ep \ep{'^*} \rangle = \int 
d{\mathbf{T}} d{\mathbf{T}}  \bar{\ep}(
{\mathbf{T}}) \bar{\ep}({\mathbf{T}}{'}) 
{\cal{P}}({\mathbf{T}}, {\mathbf{T}}{'})= 
F({\mathbf{C}}({\mathbf{r}}_{ij})),
\end{eqnarray}
where ${\mathbf{C}}({\mathbf{r}}_{ij})$ is the correlation matrix of
the shear tensor, which will be Gaussian distributed if the underlying
fluctuations are Gaussian.  The ellipticity correlation is now only a
function of the separation ${\mathbf{r}}_{ij}$.  In Section
4, we calculate the correlations of the shear tensor as well as the
moments required to find $\langle \ep \ep{'} \rangle$.  Later in
Section 4, we also examine how the correlations of the shear change if
they are sampled only at peaks of the density.  This is to account for
the fact that we are sampling galaxies, which do not form at random
positions in space.

Till now, the ellipticity correlations we have been considering are in
three dimensions.  These correlations must be projected into two
dimensions to compare with weak lensing predictions and measurements.
In Section 5, we do this projection using Limber's equation (Limber
1953).  This allows us to take into account the clustering of galaxies
whose ellipticities are sampled.


In Section 6, we examine the implications of our results for weak
lensing observations and the prospects for measuring the intrinsic
signal in on-going surveys like the SDSS and 2dF.  We conclude in
Section 7, with a more detailed discussion of our assumptions and the
uncertainties involved in our calculations.

\section{Intrinsic shape distributions}

In this section, we relate the observed, projected shapes of galaxies
to their three-dimensional shapes in the absence of lensing.  When
calculating intrinsic shape correlations, it is important to take into
account the distribution of three dimensional shapes since the
strength of the signal depends strongly on it.  For example, a
spherical galaxy will appear round when viewed from any angle,
consequently its presence will tend to suppress intrinsic shape
correlations.  We first consider the simplest case, where the galaxies
are modeled as thin disks, with the angular momentum vector
perpendicular to the disk plane.  This is a fairly good approximation
for spiral galaxies.  We then consider the effects of projecting more
realistic galaxy shapes, modeling them to be tri-axial with Gaussian
distributed axis ratios.

For a galaxy with a thin disk, the exact dependence of the observed
ellipticity is easy to calculate.  The shape of such a disk-like
galaxy depends strongly on the observing angle, appearing round
($\epsilon = 0$) when viewed face on and very elongated ($\epsilon =
1$) when viewed edge on.  When viewed from an angle $\theta$ with
respect to the perpendicular, the disk will be foreshortened by a
factor of $\cos \theta$ in one direction. The magnitude of the
observed ellipticity is then,
\begin{eqnarray}
|\epsilon| = {1 - \cos^2 \theta \over 1 + \cos^2 \theta} = 
{{1 - \hat{L}_z^2} \over {1 + \hat{L}_z^2}},   
\end{eqnarray}
where ${\mathbf{\hat{L}}} =(\sin \theta \cos \phi,\sin \theta \sin
\phi,\cos \theta),$ is the unit spin vector. The observed ellipse has
its long axis oriented perpendicular to the projected angular momentum
vector, so that $\psi = \phi + \pi/2.$

For realistic galaxies, however, the relation between the observed and
intrinsic shapes can be much more complicated. The finite thickness of
the galaxy puts an upper limit on how elongated the observed shape can
be.  Observed galaxy samples also contain a mix of morphological types
-- ellipticals, spirals and spheroidals -- each of which has a
different distribution of intrinsic shapes.

We consider the intrinsic shape distributions found by Lambas, Maddox
\& Loveday (1992; LBL hereafter), extracted from the Bright Galaxy
Survey of the APM catalog. They used triaxial models to describe the
observed ellipticity distributions of the various morphological
classes and obtained fits for the distribution of the underlying axes
ratios.  They assumed Gaussian distributions for the scaled
axes ratios ($a =1$) for all three morphological classes of the form,
\begin{eqnarray}
{\cal{P}}(b) \propto \exp \left[\frac{-(b-b_0)^2}{2 \sigma_b^2}\right],\,\,\,
{\cal{P}}(c) \propto \exp \left[\frac{-(c-c_0)^2}{2 \sigma_c^2}\right], 
\end{eqnarray}   
truncated such that $0 < c < b < 1$.  For the spiral population, they
found it necessary to include the effects of a finite disk thickness
in order to explain the deficit in the high ellipticity tail of the
observed distribution. The best-fit parameters for spirals were found
to be $b_0 = 1.0, \sigma_b = 0.13, c_0 = 0.25$ and $\sigma_c = 0.12$.
LML also demonstrated that simple oblate or prolate models were not
capable of reproducing the observations for elliptical galaxies and
that triaxial models were required. For ellipticals, the best-fit
parameters were found to be $b_0 = 0.95, \sigma_b = 0.35, c_0 = 0.55 $
and $ \sigma_c = 0.2$.  Finally, the best-fit parameters for
spheroidals were found to be $b_0 = 1.0, \sigma_b = 0.3, c_0 = 0.59$
and $\sigma_c = 0.24$.

Stark (1977) derived the relation between the three-dimensional axes
ratios and the ellipticity, which we adapt to the case at hand.
Knowing the distribution of galaxy shapes for a given galaxy type, we
can calculate the average ellipticity of a galaxy with angular
momentum at an angle $\theta$ with respect to the line of sight,
\begin{equation} 
\bar{\ep}({\mathbf{\hat{L}}}) = \int d{\mathbf{S}}_i \, 
\ep({\mathbf{S}}_i,{\mathbf{\hat{L}}}_i) \, {\cal{P}}({\mathbf{S}}_i) 
= \int_0^{2\pi} d\phi \int_0^1 db\, {\cal{P}}(b) \int_0^b dc\, {\cal{P}}(c)\, 
\ep(\theta, \phi, b, c),  
\end{equation} 
where $\ep(\theta, \phi, b, c)$ is complex and given in Appendix A.
We have performed these integrations numerically and display the
results for the amplitude in Figure 1.  By symmetry, the average
orientation angle is perpendicular to the projected angular momentum.

\begin{figure}
\centerline{\psfig{file=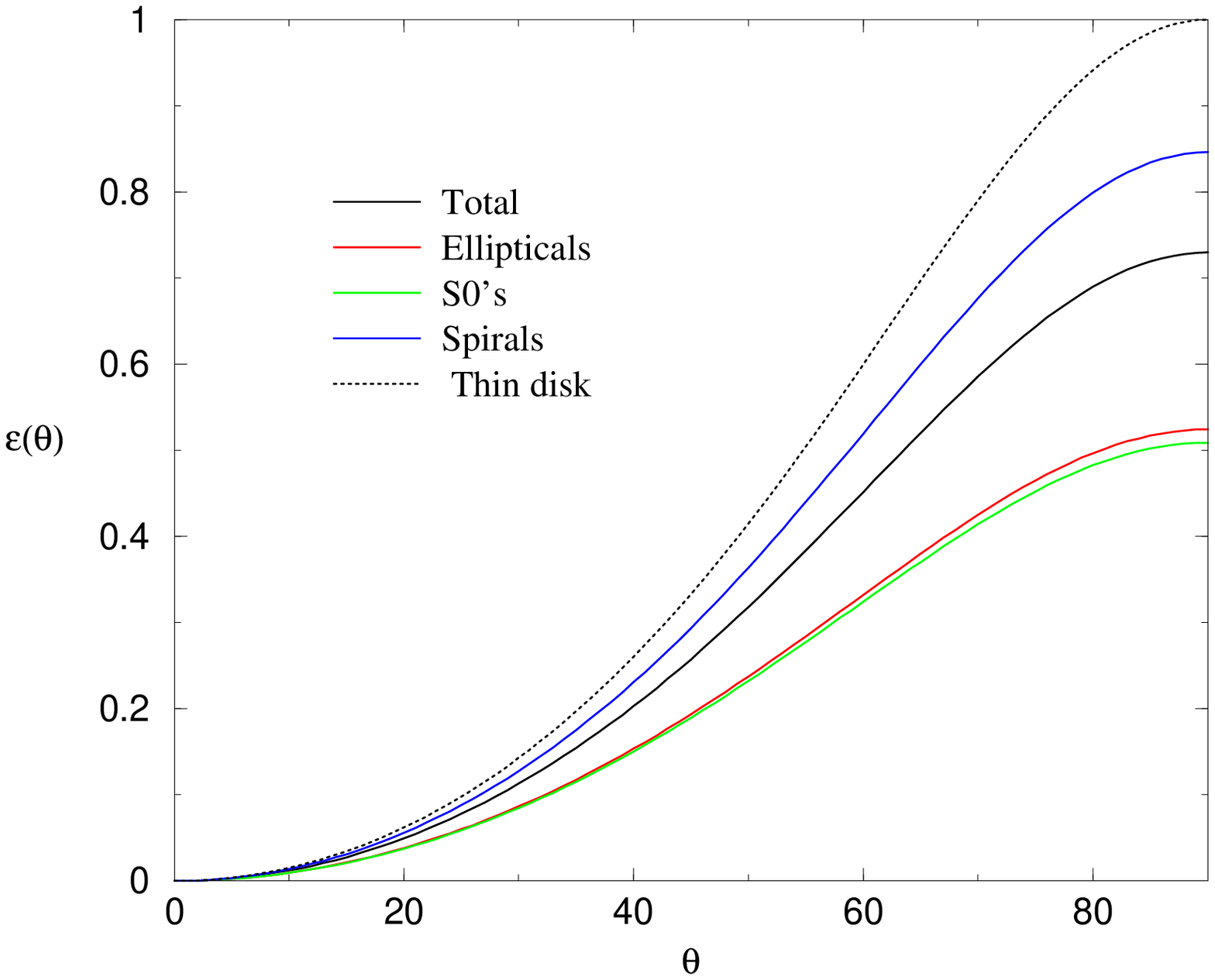,height=3.0in}
\psfig{file=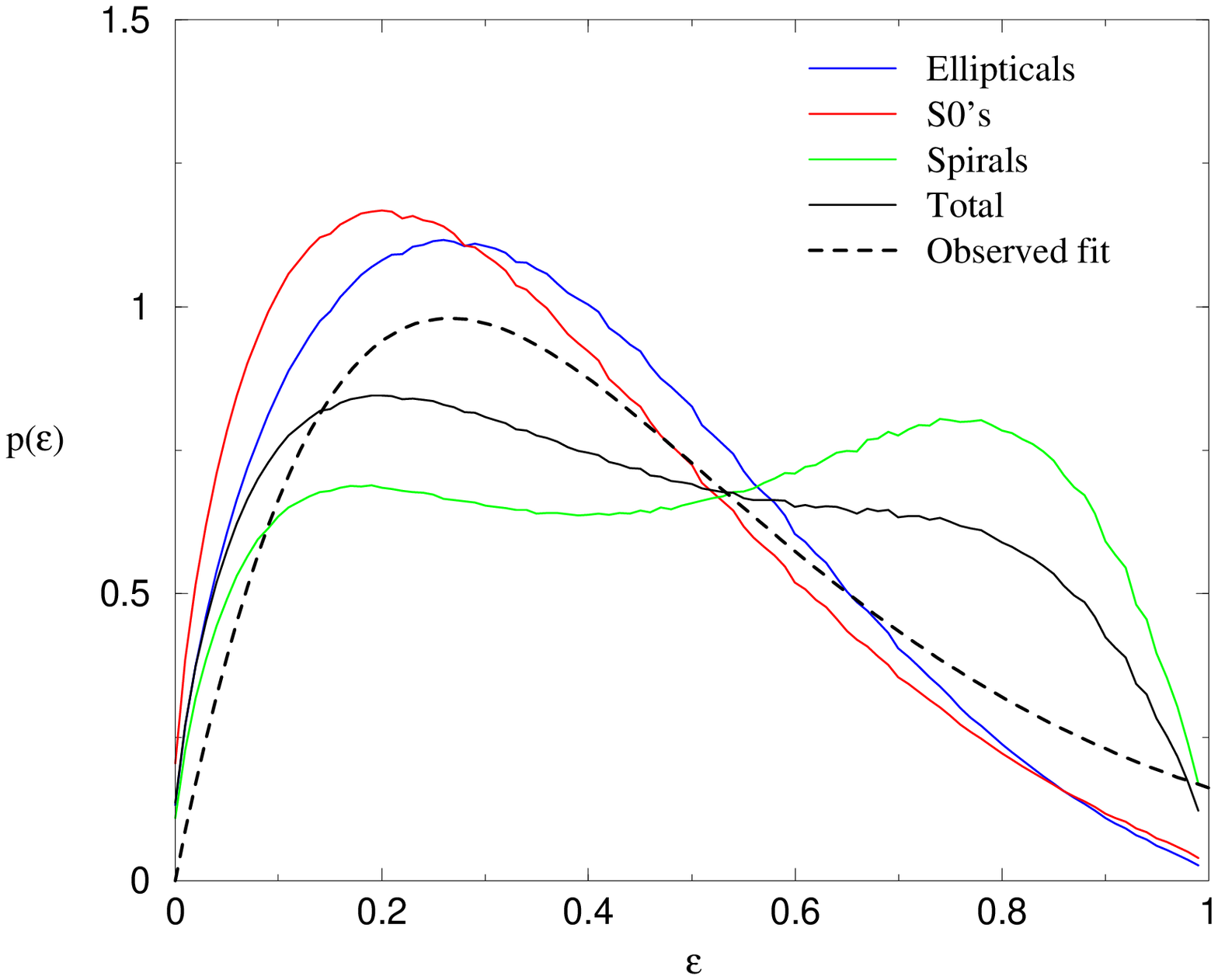,height=3.0in}}
\caption{Left panel: The average ellipticities of the different
morphological types of galaxies seen from a given angle $\theta$ with
respect to the angular momentum vector. Also shown is the result for a
sample with all types weighted by the observed fractions in the APM
survey.  All scale roughly as the thin disk case, Right panel: The
distribution of ellipticities for the different morphological types
based on the LBL intrinsic shape distributions. The differences
between the LBL model and the fit from lensing surveys (dashed curve)
could either be due to there being a different morphological mix in
high redshift surveys or evolution in the intrinsic shape
distributions, particularly for the spirals.  }
\label{fig-shapes}
\end{figure}

As can be seen in the left panel of Figure 1, $\bar{\ep}$ is different
for each morphological type. The maxima, corresponding to when the
galaxies are seen edge on, are significantly less than 1 due to the
finite thicknesses of the galaxies.  Interestingly, however, the
dependence of $\bar{\ep}$ on $\theta$ for each of the morphological
classes roughly scales identically to the disk case.  That is, it is a
good approximation to assume the same functional dependence on angle
with an overall scaling:
\begin{eqnarray}
|\bar{\epsilon} (\hat{L}_z)| = \alpha 
{{1 - \hat{L}_z^2} \over {1 + \hat{L}_z^2}}, 
\end{eqnarray}
where $ 0 < \alpha < 1 $ is a measure of the relative galaxy
thickness.  For the more realistic distributions, $\alpha$ ranges from
0.85 for spirals, down to around 0.5 for spheroids and ellipticals.

The mean redshift of the APM survey is $z_m = 0.1$ and the composition
of the sample is roughly 10\% ellipticals, 25\% spheroidals and 65\%
spirals. While these fractions might be representative of a local
field sample, the redshift distribution of background galaxies of
interest in lensing studies is considerably higher, so the
morphological mix could be much different.  To examine if this is the
case, we plot both the distribution of ellipticities for the LBL
populations and that measured for lensing studies
\begin{eqnarray}
{\cal{P}}(|\ep|) \propto |\ep| e^{-(|\ep|/0.3)^{1.15}}
\label{eqn-p_e} 
\end{eqnarray}
(Brainerd et al. 1996; Ebbels et al. 2000), which is represented by
the dashed curve on the right panel.  The LBL distributions provide a
poor fit to the results of the higher redshift field surveys. This
could be due to two effects: either a different morphological mix or a
different, perhaps non-Gaussian, distribution of intrinsic axis ratios
for spirals. In addition, since the higher redshift surveys are more
likely to be dominated by irregulars, the intrinsic shape
distributions are expected to evolve.

Though we find that the LBL distributions on the whole provide a
poor-fit to that from lensing studies, the inferred mean value for
$|\bar{\ep}|$ is in good agreement. For a thin disk, $|\bar{\ep}|_{TD}
= \pi/2 -1$, so our simplified model has $|\bar{\ep}| = \alpha
|\bar{\ep}|_{TD}= 0.57 \alpha.$ The mean of the measured ellipticity
distribution (Eqn. \ref{eqn-p_e}) is 0.42, implying that $\alpha =
0.73$. This is consistent with the mean value for $\alpha$ computed
from the APM sample on the left panel of Figure 1.  In the following
sections, we will be explicitly computing correlations between the
components of $\hat{\mathbf{L}}$ in order to calculate $\langle \ep
\ep'\rangle$ based on the definition in Eqn. (15).

\section{Ellipticity and the tidal field}

We wish to relate the ellipticity directly to the tidal field.  The
easiest way to do so is to consider the real and imaginary pieces of
the distortion field separately:
\begin{eqnarray} 
\bar{\epsilon}_+ &=& |\bar{\epsilon}| \cos(2\phi) = 
\alpha {{1 - \hat{L}_z^2} \over {1 + \hat{L}_z^2}} {{\hat{L}_y^2 -\hat{L}_x^2}
\over {\hat{L}_y^2 + \hat{L}_x^2}} 
= \alpha {{\hat{L}_y^2 -\hat{L}_x^2} \over {1 + \hat{L}_z^2}} \nonumber \\
\bar{\epsilon}_{\times} &=& |\bar{\epsilon}| \sin(2\phi) =
\alpha {{1 - \hat{L}_z^2} \over {1 + \hat{L}_z^2}} {{\hat{L}_y \hat{L}_x}
\over {\hat{L}_y^2 + \hat{L}_x^2}} = 
\alpha {{\hat{L}_y \hat{L}_x} \over {1 + \hat{L}_z^2}}. 
\label{eqn-e_L}
\end{eqnarray}
When the observation frame coincides with the frame where the stress
tensor is diagonal, $\langle \hat{L}_y \hat{L}_x \rangle = 0$ and the
expected distortion is purely real.  Next we need to relate the
angular momentum to the tidal field.

One central assumption in this work is that the expectation value of
the angular momentum at a point is solely a function of the tidal
tensor at that point. As the next step in calculating ellipticity
correlations, we need to understand this relationship more
quantitatively.  That is, we need to know the probability of a given
spin direction for a specified shear field, or effectively the form of
${\cal{P}}(\hat{\mathbf{L}}|{\mathbf{T}}).$ As discussed above, the
angular momentum of a collapsing region is given by
$L_{\alpha}\,=\,a^2(t)\,{\dot D}(t)\,\epsilon_{\alpha \beta \gamma}\,
T_{\beta \sigma}\,I_{\sigma \gamma}.$ The crucial issue is how the
moment of inertia for a collapsing object is related to the tidal
field that it experiences.  Unfortunately, this requires a precise
understanding of what determines the region which eventually collapses
into the galaxy, which in turn depends on the positions of nearby
over-densities.  This remains a major unsolved problem.

Catelan and Theuns (CT96a), studying the variance of the amplitude of
the angular momentum of galaxies, initially assumed that the tidal
tensor and the moment of inertia are entirely uncorrelated.  However,
since galaxies form preferentially at density peaks, CT96a also
consider the suppression of angular momentum that arises around peaks
in a Gaussian due to correlations between the inertia and the shear.
In both of these cases, if one considers the frame where the inertia
tensor is diagonal, the off-diagonal terms of the shear tensor are
expected to be uncorrelated with the inertia tensor.  However, around
peaks the amplitude of these off-diagonal terms is suppressed, which
results in lower angular momenta.

As we will show, the amplitude of the angular momentum has little
effect on the magnitude of ellipticity correlations, which is chiefly
determined by correlations in the directions of the spins.  For
simplicity, we will assume each component of the inertia tensor to be
Gaussian distributed, so that the resulting distribution for the
angular momentum given some shear tensor is also Gaussian distributed,
\begin{eqnarray}
{\cal{P}}({\mathbf L}|{\mathbf T}) = {1 \over (2\pi)^{3 \over 2}
|{\cal{Q}}|^{1 \over 2}} e^{- L_\alpha {\cal{Q}}_{\alpha \beta}^{-1}
L_\beta/2} 
\end{eqnarray}
where ${\cal{Q}}_{\alpha \beta} \equiv \langle L_\alpha L_\beta \rangle$ 
is the correlation matrix.
In the frame where
the shear is diagonal, the 
inertia tensor is uncorrelated with it  and  
the correlation matrix has the form, 
\begin{eqnarray}
\langle L_\alpha L_\beta \rangle \propto \left[ \begin{array}{ccc} 
(T_{33} - T_{22})^2 & 0 & 0 \\ 0 & (T_{11} - T_{33})^2 & 0 \\ 
0 & 0 & (T_{22} - T_{11})^2 \end{array} \right]. 
\end{eqnarray}
This has the same form in the peaks case studied by CT96a.
The relation can be rewritten as
\begin{eqnarray}
\langle L_\alpha L_\beta \rangle = 
\langle L^2 \rangle \left({2 \over 3} \delta_{\alpha \beta} - 
\hat{T}_{\alpha \gamma} \hat{T}_{\gamma \beta} \right), 
\end{eqnarray}
where $\hat{T}$ is the unit normalized traceless tidal tensor
($\hat{T}_{\alpha \beta} \hat{T}_{\alpha \beta} = 1$).  Note that the
angular momentum is independent of the trace of the tidal tensor, so
we can consider $\hat{T}$ to be traceless without loss of generality.  Our
final expression for the ellipticity correlations will be independent
of the proportionality factor $ \langle L^2 \rangle$, so uncertainties
in this factor will be irrelevant here.

It has been argued that the approximations made by CT96a underestimate
the correlations between the moment of inertia and the tidal field,
and result in overestimating the angular momentum that is produced
when compared to simulations.  Lee and Pen (LP00) suggest that the
CT96 approximations may do well in predicting the direction of the
angular momentum, but not its amplitude.  They consider the most
general correlation between the shear and inertia tensors,
\begin{eqnarray} 
\langle {L}_\alpha {L}_\beta \rangle = \langle L^2 \rangle \left(
{{1+a} \over 3}\delta_{\alpha \beta} 
- a \hat{T}_{\alpha \gamma} \hat{T}_{\gamma \beta}\right),
\end{eqnarray}
where $ 0 \leq a \leq 1.$ The CT96a case of uncorrelated moment of
inertia corresponds to $a=1$.  In the extreme $a=0$ case, the
direction of the angular momentum is random, independent of the tidal
tensor.  Ironically, stronger correlations between the moment of
inertia and the tidal field make the direction of the expected angular
momentum more, not less, random.  The directions can be further
perturbed by non-linear interactions, particularly if the magnitude of
the angular momentum is small originally.  LP00 investigate this in
N-body simulations and find that the relation is best fit by $a=0.24$.

The distribution of the direction of the angular momentum vector is
given by integrating over the amplitude of the momentum,
\begin{eqnarray}
{\cal{P}}(\hat{\mathbf L}|{\mathbf T})\,&=&\,\int\,L^2\,dL\,
{\cal{P}}({\mathbf L}|{\mathbf T}) \\ \nonumber &=& 
\int_0^{\infty}\,L^2\,dL\,{1 \over (2\pi)^{3 \over 2}|{\cal{Q}}|^{1
\over 2}}\,e^{- L^2\,\hat{L}_\alpha\,{\cal{Q}}_{\alpha \beta}^{-1}
\,\hat{L}_\beta/2} \nonumber \\ &=& {1 \over 4\pi |\hat{\cal{Q}}|^{1 \over 2}} 
(\hat{L}_\alpha \hat{\cal{Q}}_{\alpha \beta}^{-1} \hat{L}_\beta)^
{-{3 \over 2}}, 
\end{eqnarray} 
where $\hat{\cal{Q}} = {\cal{Q}}/\langle L^2 \rangle.$
The variance of the two point expectation value of the direction of 
the angular momentum is then 
\begin{eqnarray}
\langle \hat{L}_\alpha \hat{L}_\beta \rangle = \int d^2\hat{L} \, \, 
\hat{L}_\alpha \hat{L}_\beta {1 \over 4\pi |\hat{\cal{Q}}|^{1 \over 2}} 
(\hat{L}_\alpha \hat{\cal{Q}}_{\alpha \beta}^{-1} \hat{L}_\beta)^
{-{3 \over 2}}. 
\end{eqnarray}  
As shown by LP00 and in Appendix B, for small $a$ this implies that 
\begin{eqnarray}
\langle \hat{L}_{\alpha} \hat{L}_{\beta} \rangle  =  
{\frac{1}{3}}(1 - {\frac{3 a}{5}}) \delta_{\alpha \beta} + 
{\frac{3 a}{5}} \hat{T}_{\alpha \gamma} \hat{T}_{\gamma \beta}.
\end{eqnarray}

Combining the above results, we can compute the average ellipticity
for a given shear tensor at one point,
\begin{eqnarray} 
\bar{\ep}({\mathbf{T}}) &=& \int d{\mathbf{\hat{L}}} \,
\ep({\mathbf{\hat{L}}}) \, {\cal{P}}({\mathbf{\hat{L}}}|
{\mathbf{T}}) \nonumber \\
&=& \alpha \int d^2{{\hat{L}}}\,\,{\frac{\sin^2\theta}{1 + \cos^2 \theta}}\,[
\cos 2\phi + i \sin 2\phi]\,{1 \over 4\pi |\hat{\cal{Q}}|^{1 \over 2}}\,
(\hat{L}_\alpha \hat{\cal{Q}}_{\alpha \beta}^{-1} \hat{L}_\beta)^
{-{3 \over 2}}.
\end{eqnarray}
For small $a$, we can approximate $\hat{\cal{Q}}^{-1} \simeq {1 \over 3} 
[\delta_{\alpha \beta} - a (\delta_{\alpha \beta} - 3 \hat{T}_{\alpha \gamma}
\hat{T}_{\gamma \beta})]$.  Therefore, 
$(\hat{L}_\alpha \hat{\cal{Q}}_{\alpha \beta}^{-1} \hat{L}_\beta)^
{-{3 \over 2}} \simeq  |\hat{\cal{Q}}|^{1 \over 2}
[1 + {3 a \over 2} - {9 a \over 2} \hat{L}_\alpha \hat{L}_\beta
 \hat{T}_{\alpha \gamma}\hat{T}_{\gamma \beta}] $. 
Inserting this in the integral, by symmetry the surviving terms are  
\begin{eqnarray}
\bar{\ep}({\mathbf{T}}) &=&
{-9a \alpha\over 8 \pi} \int_0^{\pi} d\theta\,\frac{\sin^5 \theta}
{1 + \cos^2 \theta}
\int_0^{2\pi} d\phi (\cos 2\phi + i \sin 2\phi)\,\left[ 
\hat{T}_{1 \gamma}\hat{T}_{\gamma 1} \cos^2 \phi +  
\hat{T}_{2 \gamma}\hat{T}_{\gamma 2} \sin^2 \phi +  
 2\hat{T}_{1 \gamma}\hat{T}_{\gamma 2} \sin \phi 
\cos \phi \right] \nonumber \\ 
&=&
{-9a\alpha \over 8 \pi} \int_0^{\pi} d\theta\,\frac{\sin^5 \theta}
{1 + \cos^2 \theta} \int_0^{2\pi} d\phi(\cos 2\phi + i \sin
2\phi)\,\left[\,A\,+\,B \cos 2\phi\,+\,C \sin 2\phi\,\right]    
\end{eqnarray}
where $A = {1 \over 2}
(\hat{T}_{1 \gamma}\hat{T}_{\gamma 1} + \hat{T}_{2 \gamma}\hat{T}_{\gamma 2})$,
$B = {1 \over 2}
(\hat{T}_{1 \gamma}\hat{T}_{\gamma 1} - \hat{T}_{2 \gamma}\hat{T}_{\gamma 2})$ 
and $C = \hat{T}_{1 \gamma}\hat{T}_{\gamma 2}.$ 
Finally, using $ \int_0^{\pi/2} d\theta\,\frac{\sin^5 \theta}{1 + \cos^2 \theta}\,=\, 
\pi - 8/3$, we find that the integral of Eqn. (26) evaluates to 
\begin{eqnarray}
\bar{\ep}({\mathbf{T}}) &=& \bar{\ep}_+({\mathbf{T}}) + i\bar{\ep}_{\times}({\mathbf{T}})
 = a \alpha (6 -{9 \pi \over 4})\,[\,{1 \over 2}
(\hat{T}_{1 \gamma}\hat{T}_{\gamma 1} - \hat{T}_{2 \gamma}\hat{T}_{\gamma 2})
\,+\,i\hat{T}_{1 \gamma}\hat{T}_{\gamma 2}\,].   
\end{eqnarray}
The numerical factor, ${9 \pi \over 4} -6$, is very nearly unity and
we shall drop it for convenience here.  Thus, the average ellipticity
for a given tidal field is quadratic in the tidal field and is
suppressed both by a factor due to the finite thicknesses of the
galaxies ($\alpha$) and by the randomization of the angular momentum
vector ($a$).

\section{Correlations in the Tidal Field }

In the previous section we related the ellipticity to the tidal field,
and here we calculate the moments of the tidal field required to
derive ellipticity correlations.  Since the tidal tensor is the second
derivative of the gravitational potential, its statistical properties
are directly related to those of the matter density.  We assume that
the underlying density field is Gaussian distributed, which implies
that the tidal field is also Gaussian.  However, the unit normalized
tidal field which is of relevance to us will not be Gaussian
distributed.

Knowing the full statistics of the tidal field, it is possible to
calculate expectation values of observables such as the ellipticity
correlation $\langle \ep \ep{'^*} \rangle = \int d{\mathbf{T}}
d{\mathbf{T}}' \bar{\ep}( {\mathbf{T}}) \bar{\ep}^*({\mathbf{T}}')
{\cal{P}}({\mathbf{T}}, {\mathbf{T}}').$ Since the ellipticity is
quadratic in $\hat{\mathbf{T}}$, we need to compute linear and
quadratic two point functions of the normalized shear field, $\langle
\hat{\mathbf{T}}\hat{\mathbf{T}}'\rangle$ and $\langle
\hat{\mathbf{T}}\hat{\mathbf{T}}\hat{\mathbf{T}}'\hat{\mathbf{T}}'
\rangle$ respectively.  We first compute the correlations of
${\mathbf{T}}$ which are necessary to evaluate these moments.

\subsection{Correlations of ${\mathbf{T}}$}

To begin, we compute the two point expectation value of the tidal
tensor in terms of the power spectrum of fluctuations.  A Fourier
expansion of the shear tensor yields,
\begin{eqnarray}
T_{\alpha \beta}({\mathbf x})
\,=\,-\int\,d^3{\mathbf k}\,k_{\alpha}\,k_{\beta}\,\Psi 
({\mathbf k})\,e^{i {\mathbf k} \cdot{\mathbf x}},
\end{eqnarray}
where $\Psi({\mathbf k})$ is the Fourier transform of the
gravitational potential, which has a power spectrum defined as,
\begin{eqnarray}
\langle{\Psi(\mathbf k)\,\Psi(\mathbf k')}\rangle\,=\, 
\delta_{\rm D}({\mathbf k}-{\mathbf k'})\,P_{\Psi}(k).
\end{eqnarray}
>From this, it is straight forward to calculate the two-point
correlation function of the tidal tensor,
\begin{eqnarray}
C_{\alpha \beta\gamma \sigma}({\mathbf r}) &\equiv& 
\langle T_{\alpha \beta}({\mathbf x})\,T_{\gamma \sigma}({\mathbf
x^{'}}) \rangle\,=\,\int\,d^3{\mathbf
k}\,k_{\alpha}\,k_{\beta}\,k_{\gamma}\,k_{\sigma}\,P_{\Psi}(k)\, 
e^{i {\mathbf k} \cdot{\mathbf r}}, 
\end{eqnarray}
where the separation is ${\mathbf r} = {\mathbf x} - {\mathbf
x'}$.\footnote{Note that the positional vectors are in Lagrangian space 
and were denoted by $\mathbf{q}$ in Section 1.}

To evaluate the correlation function, it is useful to relate Fourier
space components back to the real space derivatives via $i k_{\alpha}
\equiv {\partial_{\alpha}}.$ Thus, we have
\begin{eqnarray}
C_{\alpha \beta\gamma \sigma}({\mathbf r}) 
\,&=&\,\partial_{\alpha} \partial_{\beta}
\partial_{\gamma} \partial_{\sigma} \int\,d^3{\mathbf
k}\,P_{\Psi}(k)\,e^{i {\mathbf k} \cdot{\mathbf r}} \nonumber \\
&=&\, 2\pi \partial_{\alpha} \partial_{\beta}
\partial_{\gamma} \partial_{\sigma}  \int \,d
k\,k^2\,P_{\Psi}(k)\,j_0(kr).
\end{eqnarray}
Here we have performed the integration over the angular directions of
the Fourier modes, and $j_0(kr) = \sin kr/kr$ is the zeroth order
spherical Bessel function.

It is useful to rewrite the derivatives as ${\partial_{\alpha}} =
({dr/dx_{\alpha}})({d/dr}) = x_{\alpha} D$, where the operator $D
\equiv (1/r)(d/dr)$.  Using the identity ${\partial_{\alpha}}
x_{\beta} = \delta_{\alpha \beta},$ we find
\begin{eqnarray}
C_{\alpha \beta\gamma \sigma}({\mathbf r}) 
&=&\,2 \pi \partial_{\alpha}
\partial_{\beta} \int\,dk\,k^2\,P_{\Psi}(k)\,(\delta_{\gamma
\sigma}\,D [j_0(kr)]\,+\,x_{\gamma}\,x_{\sigma}\,D^2 [j_0(kr)]) 
\nonumber \\ &=&\,2\pi [\delta_{\alpha \beta}\delta_{\gamma \sigma}\,
+\,\delta_{\alpha \gamma}
\delta_{\beta \sigma}\,+\,\delta_{\alpha \sigma} \delta_{\beta \gamma}] 
\int\,dk\,k^2\,P_{\Psi}(k)\,D^2 [j_0(kr)] \nonumber \\
&+&\,2\pi [r_{\alpha} r_{\beta} \delta_{\gamma \sigma}+
r_{\alpha} r_{\gamma} \delta_{\beta \sigma}+r_{\alpha} 
r_{\sigma} \delta_{\beta \gamma}+r_{\beta} r_{\gamma}
\delta_{\alpha \sigma}
+r_{\beta} r_{\sigma} \delta_{\alpha \gamma} +r_{\gamma}
r_{\sigma} 
\delta_{\alpha \beta} ]
 \int\,dk\,k^2\,P_{\Psi}(k)\,D^3 [j_0(kr)] \nonumber \\ &&+\, 2\pi
r_{\alpha} r_{\beta} r_{\gamma}
r_{\sigma}\, \int\,dk\,k^2\,P_{\Psi}(k)\,D^4 [j_0(kr)].
\end{eqnarray}
It is possible to use Poisson's equation to substitute the power
spectrum of the potential with that of the density, $k^4 P_{\Psi}(k) =
P_{\delta}(k)$, in units where $4\pi G \rho_0 =1$ and $G$ is the
gravitational constant.  Using the identity $D^{n}\,j_0(r) =
(-1)^n\,r^{-n}\,j_n(r)$, the above simplifies to
\begin{eqnarray}
C_{\alpha \beta\gamma \sigma}({\mathbf r}) & = &  
[\delta_{\alpha \beta}\delta_{\gamma \sigma}
+\delta_{\alpha \gamma} \delta_{\beta \sigma}+\delta_{\alpha
\sigma} \delta_{\beta \gamma}] 
\zeta_2(r) + \hat{r}_{\alpha} \hat{r}_{\beta} \hat{r}_{\gamma}
\hat{r}_{\sigma}\,\zeta_4(r)
\nonumber \\
&+&[\hat{r}_{\alpha} \hat{r}_{\beta} \delta_{\gamma \sigma}\,+
\hat{r}_{\alpha} \hat{r}_{\gamma} \delta_{\beta \sigma}\,+\hat{r}_{\alpha} 
\hat{r}_{\sigma} \delta_{\beta \gamma}\,+\hat{r}_{\beta} \hat{r}_{\gamma}
\delta_{\alpha \sigma}
+\hat{r}_{\beta} \hat{r}_{\sigma} \delta_{\alpha \gamma} 
+\hat{r}_{\gamma}\hat{r}_{\sigma} 
\delta_{\alpha \beta}]\,\zeta_3(r),
\end{eqnarray}
where 
\begin{eqnarray}
\zeta_n(r) = (-1)^n\,\frac{2 \pi}{r^{4-n}}\int
{dk\,k^{n-2}\,j_n(kr)\,P_{\delta}(k)}, 
\end{eqnarray}
and $j_n(kr)$ is the $n^{th}$ spherical Bessel function.  This is
identical to the expression derived in LP00, wherein it was shown that
these $\zeta$ functions are related to the density correlation
function, $\xi(r) = \int dk \, k^2 P_{\delta}(k) j_0(kr)$, and
integrals of it.

If the density field is smoothed, its correlation function levels off
as $r \rightarrow 0$.  In this limit, $\zeta_3$ and $\zeta_4
\rightarrow 0$ and the above expression reduces to
\begin{eqnarray}
C_{\alpha \beta\gamma\sigma}(0) = \zeta_2(0)  
 [\delta_{\alpha \beta}\delta_{\gamma \sigma}
+\delta_{\alpha \gamma} \delta_{\beta \sigma}+\delta_{\alpha
\sigma} \delta_{\beta \gamma}].
\end{eqnarray}
Since $\zeta_2(0) = \xi(0)/15$, this corresponds precisely to the
variance of the tidal field found by CT96a (equation (38) in Appendix
A.)  The correlation function simplifies dramatically when averaging
over directions ${\hat r}$,
\begin{eqnarray}
C_{\alpha \beta\gamma
\sigma}(r) \equiv 
\frac{1}{4\pi}\,\int\,d^2{\hat r}\,C_{\alpha \beta\gamma
\sigma}({\mathbf r})\,=\,\frac{1}{15}\xi(r)\,[\delta_{\alpha
\beta}\delta_{\gamma \sigma} + \delta_{\alpha \gamma} \delta_{\beta
\sigma} + \delta_{\alpha \sigma} \delta_{\beta \gamma}].
\end{eqnarray}
This is useful when the correlation length is much smaller than the
depth of the survey.

\subsection{Correlations of $\hat{\mathbf{T}}$ }

In this sub-section, we calculate the two and four point moments of
the unit normalized traceless tidal field, $\hat{\mathbf{T}}$.  Since
${\mathbf{T}}$ is not Gaussian distributed, these moments are not
necessarily simply related.  In the next sub-section, we use these
results to derive the final ellipticity correlation.

For simplicity of notation, it is useful to treat the six degrees of
freedom in the shear tensor as a vector, which we shall denote with
capital Roman subscripts, ${\mathbf{T}} = (T_{11}, T_{22}, T_{33},
T_{12}, T_{13}, T_{23})$.  While using this notation, it is important
to remember that the shear transforms as a tensor under rotations,
rather than as a vector.  To further simplify the notation, we shall
write the shear field at a displacement $\mathbf{r}$ from the origin
as $\mathbf{T}'$.  Thus in this notation, the correlation matrix
becomes a six by six matrix, $\langle T_A T_B' \rangle =
[C_{\mathbf{r}}]_{AB}$.

Since the tidal field is Gaussian, with a two-point correlation matrix
given by $C$, we can write the expectation value for an observable
like $\hat{T}_A \hat{T}_B$ as,
\begin{eqnarray}
\langle {\hat T_{A}}\,{\hat T_{B}}'
\rangle\,= \,\int\,\frac{d^6 T d^6 T'}
{{|C|^{1/2}}{(2\,\pi)^6}}
{\hat T}_{A}\,{\hat T_{B}}'\,e^
{-\frac{1}{2}{\vec T^{T}}{C^{-1}}{\vec T}}\,
\end{eqnarray}
where ${\vec T} = ({\mathbf{T}},{\mathbf{T}}')$. The matrix $C$
has the following block diagonal form,
\begin{eqnarray}
{C}\,=\,\left[ \begin{array}{cc} 
C_0 & C_{\mathbf r}  \\ C_{\mathbf r} & C_0 
\end{array} \right], 
\end{eqnarray} 
where $C_0$ is the zero-lag correlation matrix and $C_{\mathbf r}$
contains the two point correlations.  For galaxies that are separated
by distances greater than the smoothing scale, we can assume that
$C_{\mathbf r}\,\ll\,C_0$ and expand in powers of $C_{\mathbf{r}}
C^{-1}_0$ to invert $C$ to second order:
\begin{eqnarray}
{C^{-1}}\,\simeq\,\left[ \begin{array}{cc} {C^{-1}_0 (1 + C_{\mathbf
r} C^{-1}_0 C_{\mathbf r} C^{-1}_0)} & {- C^{-1}_0 C_{\mathbf
r}C^{-1}_0} \\ {- C^{-1}_0 C_{\mathbf r} C^{-1}_0}& {C^{-1}_0 (1 +
C_{\mathbf r} C^{-1}_0 C_{\mathbf r} C^{-1}_0)}
\end{array} \right]. 
\end{eqnarray} 

We perform a Taylor expansion of the exponential,
\begin{eqnarray}
\exp[{-\frac{1}{2}{\vec T^{T}}{C^{-1}}{\vec T}}] \simeq 
[1 + T\,C^{-1}_0\,C_{\mathbf r}\,C^{-1}_0\,T' + \frac{1}{2}
(T\,C^{-1}_0\,C_{\mathbf r}\,C^{-1}_0\,T')^2 + \,\ldots \,\,]
\exp[-\frac{1}{2}
(T\,C^{-1}_0\,T+ T'\,C^{-1}_0\,T')].
\end{eqnarray}
To evaluate the linear two point function of $\hat{\mathbf{T}}$, we
must keep terms to first order in $C_{\mathbf r} C^{-1}_0$.  The
expectation value can then be written as,
\begin{eqnarray}
\langle {\hat T_{A}\,{\hat T_{B}}}'
\rangle\,&=& \,\int\,\frac{d^6 T\, d^6 T'}
{{|C_0|}{(2\,\pi)^6}}{\hat T_A}\,{\hat T_{B}}'
[T\,C^{-1}_0 C_{\mathbf r} C^{-1}_0\,T'] 
e^{{-\frac{1}{2}
(T\,C^{-1}_0\,T+ T'\,C^{-1}_0\,T')}}\,
\nonumber \\ &=& \langle 
{\hat T_A}\,{T_C}
\rangle\,[C^{-1}_0 C_{\mathbf r} C^{-1}_0]_{CD}\,\langle 
{\hat T_B'}\,{T_D}'
\rangle,
\end{eqnarray} 
where 
\begin{eqnarray}
\langle {\hat T_A}\,{T_B}
\rangle\,\equiv\,\int {\frac{d^6 T}{(2\pi)^3 |C_0|^{1/2}}}
{\hat T_A}\,{T_B}
\,e^{{-\frac{1}{2}({T\,C^{-1}_0\,T})}},
\end{eqnarray} 
is proportional to the mean magnitude of the tidal field.  This is
evaluated in Appendix C by transforming variables, ${\cal{T}}_A \equiv
R_{AA'} T_{A'}$, to a basis in which the correlation function is
proportional to the unit matrix.  Using the results derived there, the
linear two point function is shown to be,
\begin{eqnarray}
\langle {\hat T_{A}\,{\hat T_{B}}}'
\rangle\,&=& 
\frac{64}{225\pi \zeta_2(0)} [\tilde{C}_{\mathbf{r}}]_{AB}
\end{eqnarray} 
where $\tilde{C}_{\mathbf{r}}$ is the correlation function of the
traceless part of the tidal field.  Though this was evaluated in the
large separation limit, its value at zero lag is very close to the
exact result.

The quadratic two point function of $\hat{\mathbf{T}}$ is evaluated in
an analogous way, except here we must keep terms to second order in
$C_{\mathbf r} C^{-1}_0$. Making this substitution, the quadratic two
point function is
\begin{eqnarray}
\langle {\hat T_{A}}\,{\hat T_{B}} {\hat T_{C}}' {\hat T_{D}}'
\rangle\,&=& \,\int\,\frac{d^6 T d^6 T'}
{{|C|^{1/2}}{(2\,\pi)^6}}
{\hat T}_{A}\,{\hat T_{B}}{\hat T_{C}}' {\hat T_{D}}' \,e^
{-\frac{1}{2}{\vec T^{T}}{C^{-1}}{\vec T}}\, \nonumber \\ 
&=& \langle {\hat T_{A}}\,{\hat T_{B}}\rangle \langle 
{\hat T_{C}}' {\hat T_{D}}'
\rangle\,+\frac{1}{2} \langle {\hat T_{A}}\,{\hat T_{B}} {T_{E}}
{T_{F}} \rangle\, \langle {\hat T_{C}}\,{\hat T_{D}} {T_{G}}
{T_{H}} \rangle\,[C^{-1}_0\,C_{\mathbf r}\,C^{-1}_0]_{EG}
[C^{-1}_0\,C_{\mathbf r}\,C^{-1}_0]_{FH}.
\end{eqnarray}
As before, the expectation value of 
\begin{eqnarray}
\langle {\hat T}_A\,{\hat T}_B  T_C T_D
\rangle\,= \, \int {\frac{d^6 T}{(2\pi)^3 
|C_0|^{1/2}}}
{\hat { T}_{A}}\,{\hat {{T}}_{B}} { T}_{C}{ T}_{D}
\,e^{{-\frac{1}{2}({T\,C^{-1}_0\,T})}}
\end{eqnarray}
is computed in the transformed basis and is derived in Appendix C.
The final form of the quadratic two point function is then
\begin{eqnarray}
\langle {\hat T_{A}}\,{\hat {T}_{B}} {\hat {T}_{C}}' {\hat T}_{D}'
\rangle\,  &=&  \left(\frac{1}{14
\zeta_2(0)} \right)^2 \left[[\tilde{C}_{\mathbf{r}}]_{AC} 
[\tilde{{C}}_{\mathbf{r}}]_{BD} +  
[\tilde{C}_{\mathbf{r}}]_{AD}  [\tilde{C}_{\mathbf{r}}]_{BC} \right]
+\,{\rm local\, terms} , 
\label{eqn:4ptT}
\end{eqnarray} 
where $\tilde{C}_{\mathbf{r}}$ is defined as above.  The local terms
correspond to the reducible parts of this fourth order moment and do
not contribute to the ellipticity correlation.

\subsection{Correlations of the Ellipticity}

Using the results derived earlier, we are finally in a position to
calculate ellipticity correlations.  Recall that the ellipticity
correlations are given by
\begin{eqnarray}
\langle \ep \ep'^* \rangle &=& \int
d\tilde{\mathbf{T}} d\tilde{\mathbf{T}}'  \bar{\ep}(
\tilde{\mathbf{T}}) \bar{\ep}^*(\tilde{\mathbf{T}}')
{\cal{P}}(\tilde{\mathbf{T}}, \tilde{\mathbf{T}}') \nonumber \\
 &\simeq& 
a^2 \alpha^2  \langle [\,{1 \over 2}
(\hat{T}_{1 \gamma}\hat{T}_{\gamma 1} - \hat{T}_{2 \gamma}\hat{T}_{\gamma 2})
\,+\,i\hat{T}_{1 \gamma}\hat{T}_{\gamma 2}\,]
[\,{1 \over 2}
(\hat{T}_{1 \sigma}'\hat{T}_{\sigma 1}' - \hat{T}_{2\sigma}'\hat{T}_{\sigma 2}')
\,-\,i\hat{T}_{1 \sigma}'\hat{T}_{\sigma 2}'\,] \rangle.  
\end{eqnarray}
We choose the separation vector $\mathbf{r}$ to lie in the $x-z$ plane
at an angle $\theta$ from the line-of-sight which we assume to be
parallel to the $z-$axis.  This choice implies that frames in which
the ellipticities are measured lie parallel to the projected
separation.

\begin{figure}
\label{fig-3dcorr}
\centerline{\psfig{file=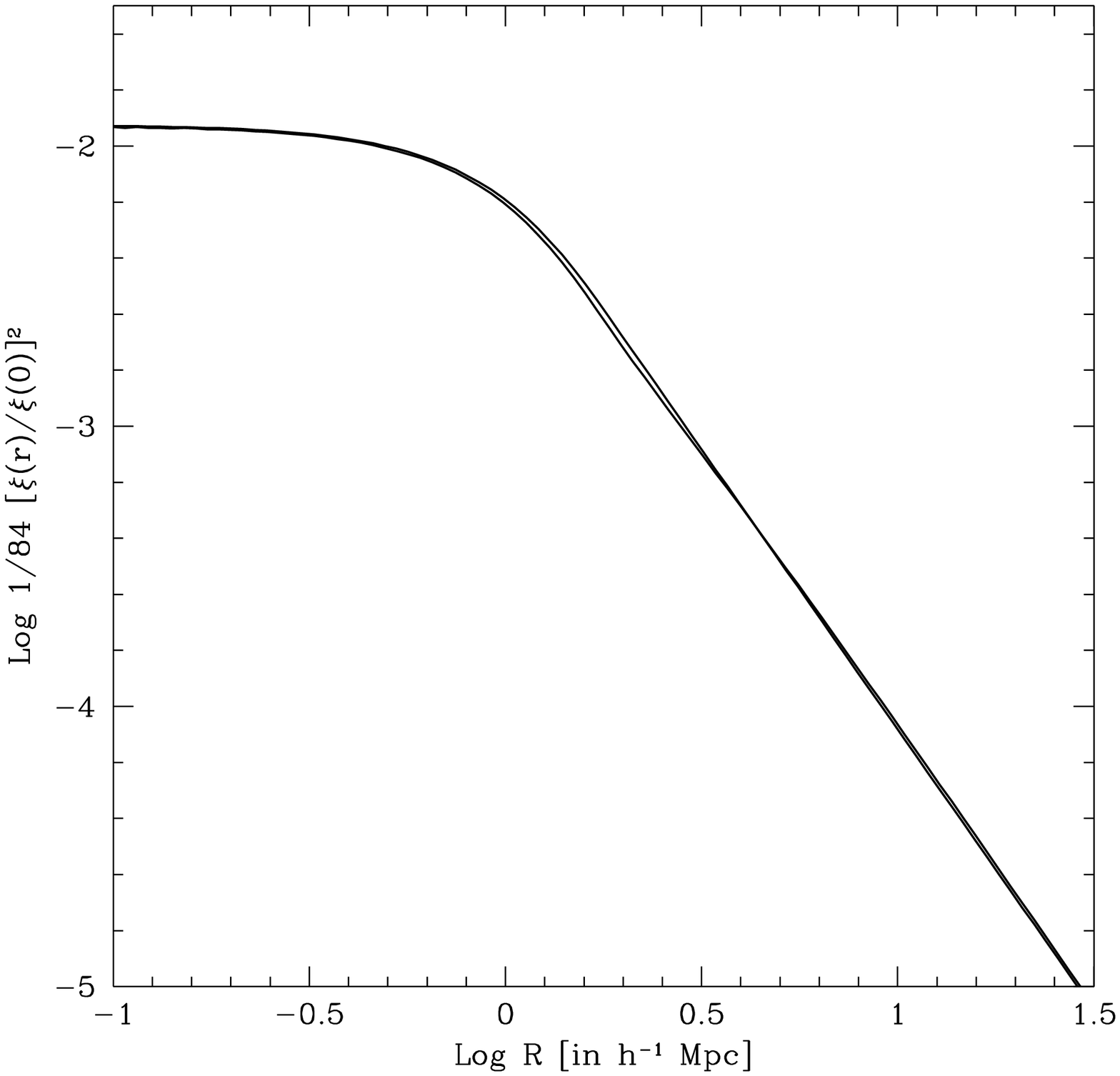,height=3.0in,width=3.0in}
\psfig{file=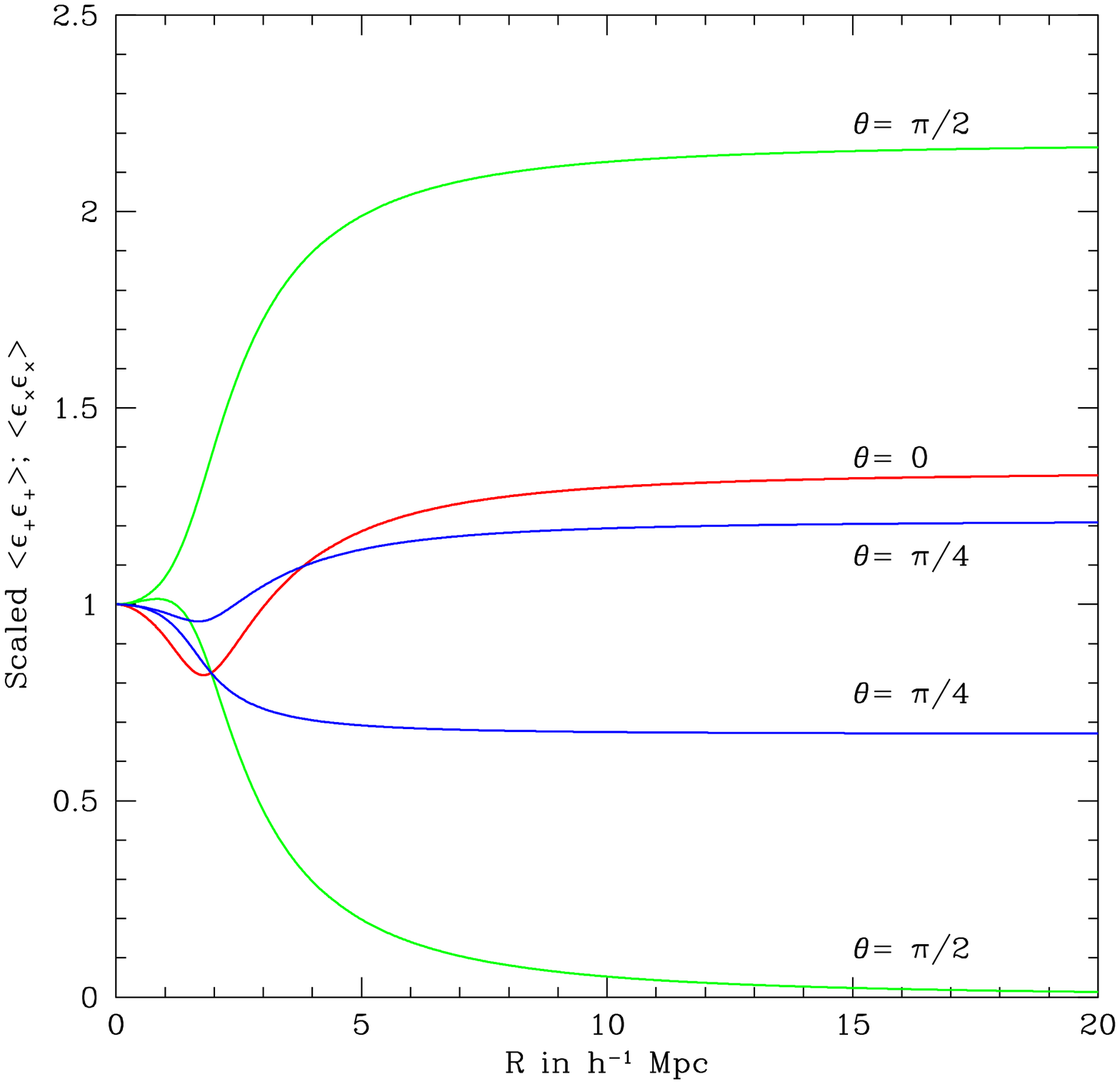,height=3.0in,width=3.0in}}
\caption{Left panel: The computed three dimensional ellipticity
correlation function averaged over angles and plotted as a function of
separation. The signal is appreciable at separations smaller than the
smoothing scale ($ < 1$ Mpc) and falls off as $\xi^2(r)$.  For
comparison we have plotted the exact derived function [solid curve] 
against the approximation of Equation (\ref{eqn:rough}) [dashed
curve]. Right panel: The
expectation value of the two components of the two point ellipticity
correlation $\langle \ep_+ \ep_+\rangle $ and $\langle\ep_{\times}
\ep_{\times}\rangle$ divided by the approximate curve (solid curve of
left panel) for various values of the viewing angle $\theta = 0,
\pi/4$ and $\pi/2$ are shown. Note that the 2 components are equal for
$\theta = 0$.}
\end{figure}

Inserting equation (\ref{eqn:4ptT}) the two non-zero components of the
ellipticity correlation are,
\begin{eqnarray}
\langle \epp \epp' \rangle =& 
\frac{a^2 \alpha^2}{144}({\frac{1}{14\zeta_2(0)}})^2
&[336\,\zeta_2^2(r) + 472\,\zeta_2(r)\,\zeta_3(r)  
   + 155\,\zeta_3^2(r) + 58\,\zeta_2(r)\,\zeta_4(r) + 
     26\,\zeta_3(r)\,\zeta_4(r)  + 3 \,\zeta_4^2(r)
       \nonumber \\  &&+ \,
      4\left(18\,\zeta_2(r)\,\zeta_3(r) -7 \,\zeta_3^2(r)
         - 8\,\zeta_3(r)\,\zeta_4(r) - \,\zeta_4^2(r) \right)
         \,\cos 2\,{\theta} \nonumber \\ &&+ 
      \,\left( 17 \,\zeta_3^2(r) + 6 \,\zeta_2(r)\,\zeta_4(r)
	 + 6 \,\zeta_3(r)\,\zeta_4(r) + \,\zeta_4^2(r)
          \right) \,\cos 4\,{\theta}],
\end{eqnarray}
and 
\begin{eqnarray}
\langle \epc \epc' \rangle =&
\frac{a^2 \alpha^2}{18} ({\frac{1}{14\zeta_2(0)}})^2
&[ 42\,\zeta_2^2(r) + 59\,\zeta_2(r)\,\zeta_3(r) +
	13\,\zeta_3^2(r) + 5\,\zeta_2(r)\,\zeta_4(r) + 
	\zeta_3(r)\,\zeta_4(r) \nonumber \\
        &&+ \left( 9\,\zeta_2(r)\,\zeta_3(r) 
	+ 5 \,\zeta_3^2(r) + 3 \,\zeta_2(r)\,\zeta_4(r) 
         - \,\zeta_3(r)\,\zeta_4(r) \right) 
       \cos 2\,{\theta} ] .
\end{eqnarray}
For a simple model with $\xi(r) \propto 1/r$, $\langle \epp \epp'
\rangle$ and $\langle \epc \epc' \rangle$ are plotted in Figure
\ref{fig-3dcorr}, computed assuming top-hat smoothing on a 1 $h^{-1}$
Mpc scale.

These functions are explicitly anisotropic and depend on the angle
between $\mathbf{r}$ and the line of sight.  Much of the angular
dependence can be understood intuitively by considering the symmetries
of the problem.  When the line of sight is parallel to $\mathbf{r}$
($\theta = 0$), there is no longer a distinction between $\epp$ and
$\epc$, so that $\langle \epp \epp' \rangle$ and $\langle \epc \epc'
\rangle$ are identical.  In addition these correlations are invariant
under the transformations $\theta \rightarrow -\theta$ and $\theta
\rightarrow \pi - \theta$ so that the only surviving terms are either
constant or proportional to $\cos 2\theta$ or $\cos 4 \theta$.  This
anisotropy is demonstrated in the right panel of Figure \ref{fig-3dcorr}.

Asymptotically at large $r$, the angle averaged behavior is approximately 
\begin{eqnarray}
\langle \ep \ep{'^*} \rangle  \simeq {a^2 \alpha^2 \over 84} {\xi^2(r) \over 
\xi^2(0)}.
\label{eqn:rough}
\end{eqnarray}
This is very close to the exact expression at large $r$, as is shown
in the left panel of Figure \ref{fig-3dcorr}.  The factor of 1/84 has
been derived for large separations.  At zero lag, it can be
computed directly from the fourth moment of the unit normalized
traceless shear tensor and is found to be 1/60.

\subsection{Peaks in a Gaussian field} 

Galaxies do not form at random positions, but at peaks of the density
field (Bardeen et al. 1986). It is possible that this sampling could
bias the expected correlation of galaxy ellipticities since our
analytic correlations have been computed for random points.  We
examine such a potential bias using numerical realizations of Gaussian
fields, and also use these to check the validity of our analytic
results.

We create realizations of Gaussian fields on a 512$^3$ grid with a
power spectrum $P_{\delta} \propto k^{-2},$ corresponding to a density
correlation which falls off as $1/r.$ The density field is smoothed with
a spherical top-hat filter of approximately four grid units.  Peaks
are identified as positions where the density field exceeds the value
at each of its six nearest neighbors. The tidal field is calculated at
each point using differencing. At each point, we compute and subtract
the trace and finally unit normalize the tidal tensor.

We checked first that the moments of the normalized tidal field match
our analytic expectations at zero lag.  For example, we can
analytically calculate the variance of one component of the tidal
tensor, $\langle \hat{T}_{xx}^2 \rangle$.  From isotropy, this can be
shown to be $2/15$, which we have verified in the realizations.  We
have also checked numerically other exact quadratic and quartic
relations such as $\langle \hat{T}_{xy}^2 \rangle = 1/10$, $\langle
\hat{T}_{xx}^4 \rangle = 4/105$ and $\langle \hat{T}_{xx}^2
\hat{T}_{xy}^2 \rangle = 1/105.$

Using the definition in Equation (48), we compute the correlation
function of the (angle averaged) ellipticity on the grid both for
peaks and random field points, assuming $a = \alpha = 1$. These are
compared with the derived analytic results in Figure 3.  The
correlation function for the peaks ($\sim$ 70,000 in the box) and the
field points are in excellent agreement for both small and large
separations. At zero lag, they both asymptote to the analytic result
of $1/60$ (marked in Figure 3 by the horizontal long-dashed line).
The maximum deviation occurs around the smoothing scale and is of the
order of 10\%.  This implies that while peaks might preferentially be
sites of galaxy formation, as far as intrinsic ellipticity
correlations are concerned, there is no substantial bias between peaks
and random field points.

The numerical correlation function starts dropping below the exact
analytic expectation (plotted as the solid curve) for large
separations. This is due to missing large-scale power on the grid: an
analytic calculation that incorporates the same lack of power on large
scales as the grid is shown as the dashed line. Its agreement with the
numerical results demonstrates that the steeper fall off of the
numerical estimates is indeed an artifact of the finite box
size. Clearly this is a worry for all numerical computations of the
ellipticity correlation function. Additionally, on separations smaller
than the smoothing scale, our analytic estimates fall below the
numerical results and the exact values derived from the one point
moments.  This discrepancy arises because the approximation we made in
computing the analytic results, $C(r)/C(0) \ll 1,$ is invalid on small
scales.

\begin{figure}
\centerline{\psfig{file=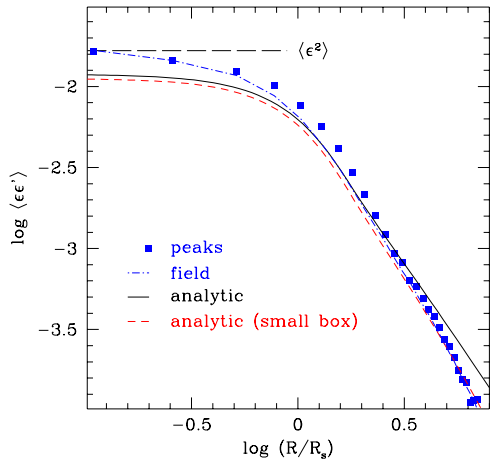,height=3.0in,width=3.0in}}
\caption{The angle averaged ellipticity correlations from Gaussian
field realizations on a $512^3$ grid compared to the analytic results.
The correlations at peaks of the density field (shown as solid
squares) appear to match well with the results from random positions
(the dot-dashed line). At zero-lag they both asymptote to the exact
result ($1/60$) shown by the horizontal dot-dashed line.  The full
analytic correlations (solid curves) as well as the analytic estimate
where the large scale power has been removed (dashed curve) to account
for the finite size of the realizations are both shown.  These are not
valid below the smoothing scale. The analytic results match well in
the region where they are valid, i. e. at scales larger than the
smoothing length, but note the importance of large-scale power (here
we have set $a=\alpha=1$).}
\label{fig-peaks}
\end{figure}

\section{The projected correlation functions} 

Up to now we have been focusing on the three dimensional ellipticity
correlation function.  While this is in principle observable (Pen et
al. 2000), weak lensing studies usually consider the ellipticity
correlation projected onto the sky. The projection of the intrinsic
signal will enable us to compare directly with weak lensing estimates
and judge its importance as a possible contaminant for these
measurements.

\subsection{Limber's equation}

We begin by making some general comments about the basis dependence of
the two dimensional correlation functions.  Previously, we calculated
the three dimensional correlation functions in a special basis, one
where the separation vector was coplanar with one of the axes used to
define the ellipticity.  We showed that the ellipticity correlations
in this basis were functions only of the 3-D separation and the angle
$\cos \theta = \hat{\mathbf{z}}\cdot\hat{\mathbf{r}}$.  In two
dimensions, this basis is equivalent to taking one axis vector to be
parallel to the 2-D separation $\vec{r}$.
\footnote{For clarity, we use the bold face for three dimensional
vectors and the notation $\vec{r}$ for 2-D vectors.}  In this special
basis, which hereafter will be denoted by the superscript $r$, the
ellipticity correlations will only be functions of the distance
between the two points.  We will denote these functions as,
\begin{eqnarray}
\xi_{+}(|\vec{r}|) \equiv \langle \epsilon^r_{+}({\vec{x}})
\epsilon^r_{+}({\vec{x}+\vec{r}}) \rangle;
\,\, \xi_{\times}(|\vec{r}|) \equiv \langle \epsilon^r_{\times}({\vec{x}})
\epsilon^r_{\times}({\vec{x}+\vec{r}}) \rangle.
\end{eqnarray}
The cross correlation,  
$\langle \epsilon^r_{+}({\vec{x}}) \epc^r({\vec{x}+\vec{r}})
\rangle,$ 
is zero due to parity ($x \rightarrow -x$) invariance. 

The basis for these correlation functions depends on the separation
vector, and thus on which pair of galaxies one is considering.  It is
often useful to work in a fixed basis on the sky for the ellipticities.
The ellipticity measured in an arbitrary basis with an angle $\phi$
relative to the separation vector is given by:
\begin{eqnarray}
\epsilon_{+}&=& \epsilon_{+}^{r} \cos 2\phi - \epsilon_{\times}^{r} \sin
2\phi; \nonumber \\
\epsilon_{\times} &=& \epsilon_{+}^{r} \sin 2\phi + \epsilon_{\times}^{r} \cos
2\phi.
\end{eqnarray}
In such a fixed basis, the correlation function depends 
on $\phi$ as, 
\begin{eqnarray}
C_1(|\vec{r}|,\phi) \equiv \langle \epsilon_{+} \epsilon_{+}' \rangle &=&
\xi_{+}(|\vec{r}|) \cos^2 2\phi \,+\,\xi_{\times}(|\vec{r}|) \sin^2 2\phi\nonumber \\
C_2(|\vec{r}|,\phi) \equiv \langle \epsilon_{\times} \epsilon_{\times}'
\rangle &=& \xi_{\times}(|\vec{r}|) \cos^2 2\phi \,+\,\xi_{+}(|\vec{r}|) \sin^2 2\phi,
\end{eqnarray}
for all pairs separated by $|\vec{r}|$ at an angle of
$\phi$ with respect to the chosen basis.  The sum of these is a
function of the separation only, while the difference has a simple
dependence on $\phi$:
\begin{eqnarray}
C_1(|\vec{r}|,\phi) + C_2(|\vec{r}|,\phi) = \xi_{+}(|\vec{r}|) + \xi_{\times}(|\vec{r}|);\,\,\,
C_1(|\vec{r}|,\phi) - C_2(|\vec{r}|,\phi) = [\xi_{+}(|\vec{r}|) - \xi_{\times}(|\vec{r}|)] \cos 4\phi.
\end{eqnarray}
Recent measurements of the shear from weak lensing have focused on the
variance of the magnitude of the ellipticity averaged over a patch,
which depends only on the sum.

We next consider the projection into two dimensions and use an
approach similar to that used by Heavens et al. (2000) and Croft \&
Metzler (2000).  Assuming that we are working on a small area of the
sky, the observed patch of sky is approximated by a plane.  The
projection uses Limber's equation to take into account the clustering
of galaxies,
\begin{eqnarray}
\xi_{+}(|\vec{r}|)\,&=&\,\frac{\int z_1^2 z_2^2 dz_1 dz_2 \psi(z_1)\,\psi(z_2)\,
[1 + \xi_{gg}(r)]\,\langle \epp(\mathbf{x_1}) \epp(\mathbf{x_2}) 
\rangle}{\int z_1^2 z_2^2 dz_1 dz_2 \psi(z_1)\,\psi(z_2)\,
[1 + \xi_{gg}(r)]} \nonumber \\ 
\xi_{\times}(|\vec{r}|)\,&=&\,\frac{\int z_1^2 z_2^2 dz_1 dz_2 \psi(z_1)\,\psi(z_2)\,
[1 + \xi_{gg}(r)]\,\langle \epc(\mathbf{x_1}) \epc(\mathbf{x_2}) 
\rangle}{\int z_1^2 z_2^2 dz_1 dz_2 \psi(z_1)\,\psi(z_2)\,
[1 + \xi_{gg}(r)]},
\label{eqn-limber}
\end{eqnarray} 
where $|\vec{r}|^2 = (r^2 - (z_1 -z_2)^2)$, $\psi(z)$ is the
observational selection function and $\xi_{gg}(r)$ is the
galaxy-galaxy correlation function.

Note that while the ellipticity correlation function is calculated in
Lagrangian coordinates, the projection is performed in Eulerian space.
The Eulerian separations will differ from the Lagrangian ones due to
peculiar velocities of galaxies, which we have ignored here.  For
galaxies near each other, the Eulerian separations will in general be
smaller than the corresponding Lagrangian ones. This will result in a
suppression of the intrinsic projected ellipticity correlations. We
expect this effect to be small at large separations, but it could be
significant closer in.

\begin{figure}
\centerline{\psfig{file=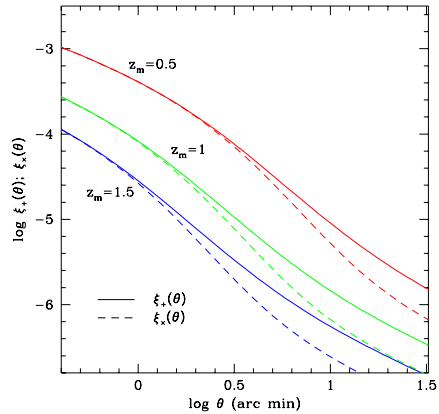,height=3.0in}}
\caption{The projected 2-d ellipticity correlation functions, 
 $\xi_+$ and $\xi_\times$,   for
various values of the median redshift of the distribution of
galaxies.  
Here we have assumed the galaxies are well described by thin disks and are 
perfectly aligned with the linear predictions ($a=\alpha=1$.) 
but deviate at larger separation. Both increase strongly with
decreasing redshift, roughly proportional to $z_m^{-2}$.}
\label{fig-2d}
\end{figure}

\subsection{Qualitative Features}

In Figure \ref{fig-2d} we plot the projected correlation function
calculated for the simple model in which the density correlation falls
off as $1/r$.  We have assumed a top-hat smoothing scale of 1
$h^{-1}$Mpc.  The clustering term is taken to be of the form $\xi_{\rm
gg} = (r/R_s)^{-\beta}$ where $R_s = 5 h^{-1}$ Mpc is the clustering
scale and $\beta = 1.8$ (Loveday et al. 1992).  Finally, following
Heavens et al., the selection function is taken to be $\psi(z) =
e^{-(z/z_0)^{1.5}}$ which has a mean redshift of $z_m = 1.4 z_0$.  In
the figure we plot the functions $\xi_+$ and $\xi_\times$ for
$a=\alpha=1$. They are nearly identical at small angular scales, but
deviate at larger separation.

The qualitative features of these correlation functions can be
understood by looking at the various scales in the problem: $R_m$, the
mean depth of the survey; $R_s$, the clustering length of galaxies and
$R_0$, the smoothing scale. If the density correlation function falls
off as $1/r$, the three dimensional ellipticity correlation scales as
$\xi_+ \simeq \xi_{\times} \simeq a^2 \alpha^2 R_0^2/ 84 (R_0^2 +
r^2),$ where we have used the approximation from
Eqn. (\ref{eqn:rough}). Both the numerator and the denominator of
Equation (\ref{eqn-limber}) contain a clustering term, which dominates
at small angular separations and a mean contribution.

The denominator in Limber's equation is essentially the average number
of galaxies within an angular distance $\theta$ from a given galaxy
out to the volume of the survey.  At large separations, the
denominator scales as the volume squared, or $R_m^6$.  Corrections
from clustering are of order $R_m^6 (R_s/R_m)^\beta \theta^{1-\beta},$
which become important at angles $\theta< (R_s/R_m)^{\beta \over \beta
- 1}.$ For a survey with median redshift $z_m \sim 1$, this occurs at
about 2 arc seconds. For a shallower survey with $z_m \sim 0.1$, this
occurs at much larger scales, of order a few arc minutes.

The numerator in Limber's equation is the projected ellipticity
correlation function, weighted by the number of pairs at a given
separation.  Again, the clustering term dominates at small scales.
For separations less than $\theta \sim R_0/R_m$, the three dimensional
correlations are effectively constant with an amplitude of $1/60$, so
the behavior is identical to that of the denominator, $a^2 \alpha^2
R_m^6 (R_s/R_m)^\beta \theta^{1-\beta}/60.$ At very large separations,
$\theta > R_s/R_m,$ clustering is not important but the numerator also
falls off inversely with angular separation, $R_0^2R_m^2 \theta^{-1}/84.$
Between these regimes, $R_0/R_m < \theta < R_s/R_m$, there is a
transition where the clustering contribution falls off quickly.

Thus the projected correlation functions have a number of distinct
regimes.  At very small separations, clustering dominates both the
numerator and the denominator, leaving the correlation constant ($a^2
\alpha^2/60$ for $\theta< (R_s/R_m)^{\beta \over \beta - 1}.$) On
slightly larger scales, but smaller than $R_0/R_m$, clustering
dominates the numerator, but not the denominator, and the correlation
falls off as a power law, $\xi_\ep \sim a^2 \alpha^2/60
(R_s/R_m)^\beta \theta^{1-\beta}.$ There is then a brief transition
region where the correlation falls fairly quickly.  Finally on very
large scales, the mean values dominate both the numerator and the
denominator and the correlation falls off as $a^2 \alpha^2\theta^{-1}
R_0^2 /(84R_m^{2}).$

It is straight forward to understand the dependence of the correlation
functions on the mean redshift of the survey.  The typical 3-d
separation of galaxies with a given angular separation is directly
proportional to the survey depth $R_m$.  Thus if the three dimensional
ellipticity correlations fall off as $r^{-n}$, then the projected
correlations fall off as $R_m^{-n}$.  For the case we have been
considering, $\xi \propto 1/r$, so that the ellipticity correlations
fall off as $1/r^2$, and the projected correlation drops as
$z_m^{-2}$.  This is clearly seen in Figure \ref{fig-2d}.

\section{Intrinsic alignments versus weak lensing}

\begin{figure}
\centerline{\psfig{file=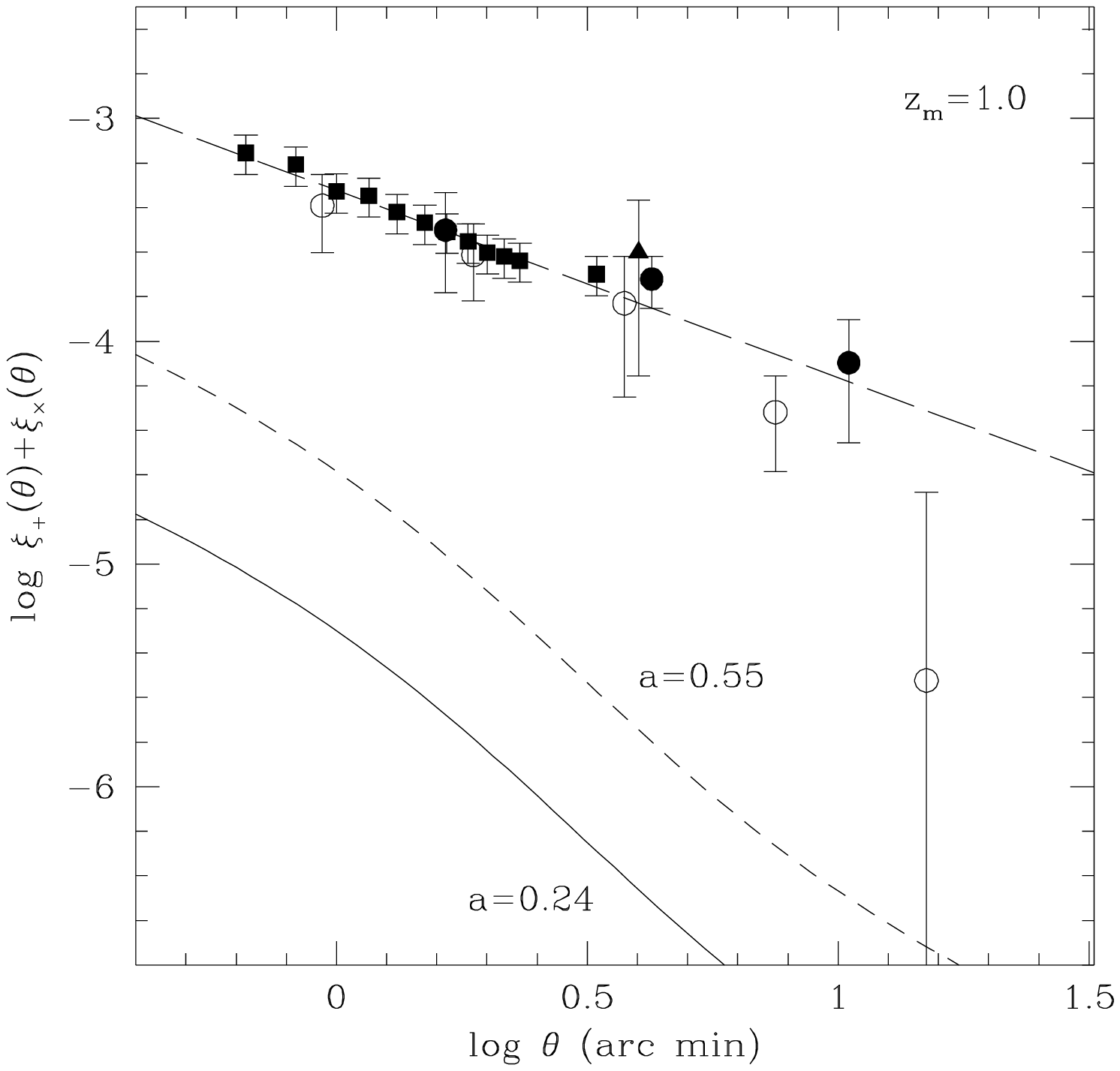,height=3.0in}
\psfig{file=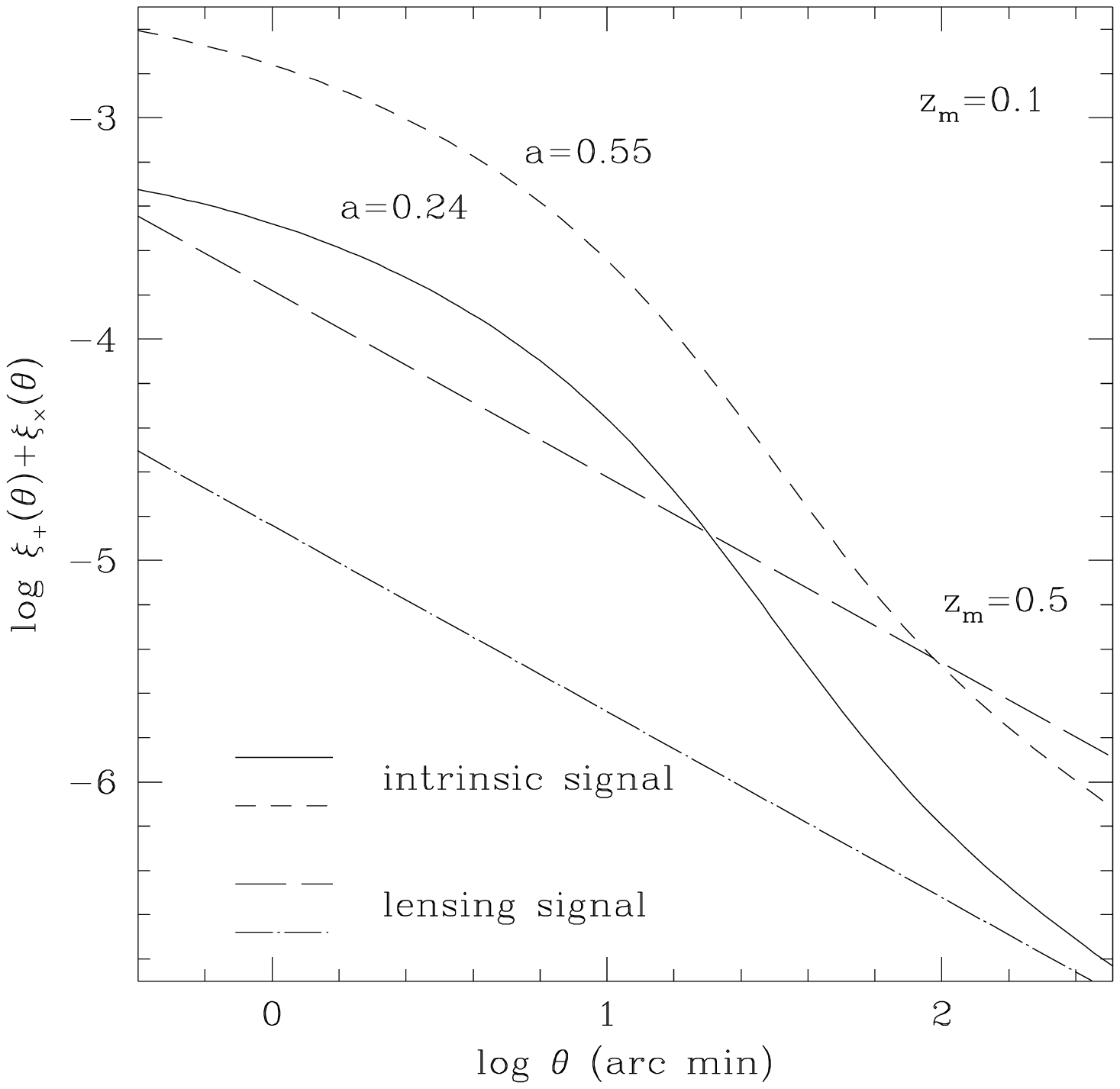,height=3.0in}}
\caption{ The intrinsic correlation signal versus the predictions from
weak lensing and current observations.  Left panel:
$\xi_+(\theta)+\xi_\times(\theta)$ for a median redshift of 1,
compared to the measured shear correlation function. At small
separations, the intrinsic signal is approximately one percent of the
measured value. The amplitude depends on the value of the assumed
average galaxy thickness ($\alpha$) and the parameter $a$ that
describes how well the angular momentum of the galaxy is correlated
with the shear field.  We plot $a=0.24$ (full line) and $a=0.55$
(short-dashed line) which correspond to the values inferred from
numerical simulations by LP00 and Heavens et al. (2000) respectively.
$\alpha=0.73$ corresponds to the value determined from the observed
distribution of ellipticities (Ebbels et al. 2000). The data are: van
Waerbeke et al. (2000) -- solid squares ; Wittman et al. (2000) --
filled circles ; Kaiser et al. (2000) -- open circles; and Bacon et
al. (2000) -- filled triangle. The long-dashed line is the theoretical
prediction from Jain \& Seljak (1997) computed for a
$\Omega_\Lambda=0.7$ galaxy cluster normalized flat universe, $\sim
4.75\times10^{-4}(\theta/{\rm arc min})^{-0.84}$.  Right panel: as in
the left panel but for the predictions for a shallower survey such as
SDSS and 2dF with median redshift $z_m=0.1$. The intrinsic signal is
again shown for two values of $a$, and the theoretical prediction for
weak lensing is the long-dashed line (for $z_m=0.1$) and
dotted-long-dashed (for $z_m=0.5$). The lensing prediction for
$z_m=0.1$ is extrapolated from the Jain \& Seljak fit beyond the
stated range of validity. For such low redshifts, the intrinsic signal
is significant and may dominate over the lensing contribution for most
scales.}
\label{fig-weak}
\end{figure}

In the previous sections we presented an analytic expression for the
intrinsic ellipticity correlation function, which we now evaluate for
realistic surveys.  We also compare this with recent measurements of
cosmic shear and with theoretical weak lensing predictions at high and
low redshifts.

The amplitude of intrinsic correlations depends on both the mean
thickness of galaxies and on their degree of alignment with the tidal
field.  In Section~2, we argued that observed shapes of galaxies are
characterized by $\alpha=0.73$.  The degree of alignment of the
galaxies with the predictions from linear theory, parameterized by
$a$, was measured by Lee and Pen (2000) in N-body simulations and
found to be fairly small, $a \simeq 0.24$.  This implies that the
correlations could be suppressed by non-linear effects.

However, we are interested in the correlations of spins with each
other, not necessarily in how they align with the predictions from
linear theory.  The LP00 measurement was a one-point measurement, and
thus can not account for `correlated randomizations.'  Non-linear
interactions between galaxies could lessen the correspondence of their
spins with the linear predictions without changing how well the spins
correlate with each other.  Thus using the one point value of $a$ will
likely underestimate the amplitude of the spin correlations.  The
measurements of the three dimensional spin correlations from the Virgo
simulations (Heavens et al. 2000) can be used to measure an effective
$a$ which takes this effect into account.  At 1 Mpc, they find a
correlation of approximately $5 \times 10^{-3}$ which is in remarkable
agreement with that found by Pen et al. observationally.  When
compared to the analytic prediction of $a^2 \alpha^2/60,$ this yields
an effective value of $a = 0.55$.  (Heavens et al. treat the galaxies
as thin disks, so that $\alpha = 1$.)  Here we will present results
for both $a=0.24$ and $a=0.55$ to demonstrate the possible uncertainty
of our predictions.

In the left panel of Figure~\ref{fig-weak}, we plot the sum of the
intrinsic correlation functions, $\xi_++\xi_\times$, for a median
redshift of 1, a galaxy smoothing scale of $1 h^{-1}$ Mpc and the
parameter choices described above.  We also show the measured shear
variance compiled from the recent literature.  The weak lensing
prediction from Jain \& Seljak (1997) for an $\Omega_\Lambda=0.7$
galaxy cluster normalized flat universe fits the data well.  At small
separations, the intrinsic signal contaminates the lensing one at the
level of a few per cent, modulo the uncertainties in $\alpha$ and $a$.

Note that there can be ambiguities in 
plotting measurements of the cosmic shear.  One issue is whether the 
correlation function or the tophat variance is plotted.  For a simple 
$1/\theta$ correlation function, the variance is nearly a factor of two 
larger than the correlation at the same scale.  In addition, some authors 
plot the variance as a function of the tophat smoothing radius, while others 
instead plot it as a function of the diameter.  Finally, 
some authors quote the
variance of each component of the complex shear field, while others
quote the variance of the modulus of $\epsilon$. 
Here we plot the theoretical predictions for 
$\xi_++\xi_\times$, the correlation 
of the modulus of the ellipticity.  In contrast, when plotting the data, 
we have used the modulus variance for a given tophat radius.  
Since the correlation function implied by the data is slightly lower than the 
variance, 
the relative contribution of the intrinsic correlations is somewhat larger 
than is naively implied by the figure. 

The relative importance of the intrinsic correlations increases
dramatically as the depth of the survey is reduced.  As the observed
galaxies are closer to us, the lensing signal falls because there is
less intervening matter to lens them, while the intrinsic signal grows
since the galaxies are physically closer to each other for a given
angular separation.  Jain \& Seljak (1997) show that the lensing
signal scales as $z_m^{1.52}$ for a flat $\Omega_{\Lambda} = 0.7$
universe.  In contrast, if the density correlation function is
proportional to $r^{n}$, then the ellipticity correlation scales as $
z_m^{2n}$.  On galaxy scales, $n\simeq -1$, hence the intrinsic
amplitude grows rapidly since the signal scales as $\sim z_m^{-2}.$

In the right panel of Figure~\ref{fig-weak} we compare the intrinsic
correlations for a shallow survey with $z_m=0.1$ such as SDSS or 2dF
to the lensing signal expected from the theoretical analysis of Jain
\& Seljak (1997). We show their fits for $z_m=0.5$ and also
extrapolate the fit to $z_m=0.1$. The latter is beyond the stated
range of validity, but should give an approximate idea of the lensing
amplitude. For low $z_m\le 0.3$, the intrinsic signal is significant
and may dominate over the lensing contribution on most
scales. Clearly, large surveys like 2dF and SDSS offer exciting
possibilities for measuring intrinsic shape correlations.

There are other important distinctions between the lensing and the
intrinsic correlation signals.  For example, the lensing signal
depends on the amplitude of the mass fluctuations, parameterized by
$\sigma_8$.  In contrast, the intrinsic correlations depend only on
correlations of the direction of the shear field and are therefore
largely independent of the amplitude of the fluctuations.

In addition, another difference arises in how the intrinsic signal
depends on morphological type.  Weak lensing is in some sense
democratic, as all galaxy types are distorted in the same way.  This
is not the case for intrinsic correlations however.  We have shown
that this signal depends on the distribution of axis ratios.  Spiral
galaxies are characterized by $\alpha \simeq 0.85$, while ellipticals
and spheroidal galaxies typically have $\alpha \simeq 0.5$.  In
addition, the alignment of the angular momentum with the shortest axis
is likely to have more scatter in elliptical galaxies, resulting in an
effective lowering of the value of $a$.  Thus we expect the intrinsic
correlation to be suppressed by more than a factor of two.  This
hypothesis can be checked observationally by using color criteria to
separate the morphological types of galaxies, since ellipticals tend
to be redder than spirals.
 
\section{Summary}

In this paper, we have presented a calculation of intrinsic
correlations in the observed ellipticities of galaxies resulting from
angular momentum couplings. We have focused on the angular momentum
which arises in linear theory and is associated with the local tidal
field. The three dimensional spin correlations were projected using
Limber's equation to obtain the expected 2-d ellipticity
correlations. These intrinsic correlations were shown to dominate over
the weak lensing signal for shallower surveys.

A number of assumptions were made in order to make the calculation
tractable. Foremost of these is the assumption that angular momentum
plays the central role in aligning the observed galaxy shapes.  Other
factors, such as the initial distribution of matter which fell in to
form the galaxy, could also conceivably have contributed to the
observed shapes. However, the angular momentum is special in that it
is approximately constant during the later evolution of the galaxies.  
Galaxies
typically have had many dynamical times to virialize, and we expect
most of the dependence on the initial matter distribution to be lost.
This is particularly true for spirals, and holds for ellipticals  
which are slow rotators and have spin parameters of the order of 10\%. 
Although their rotational time scale ($\sim 1 $ Gyr) is much longer than 
their dynamical time ($\sim 100$ Myr), they have undergone enough rotations 
in a Hubble time to 
erase any memory of the alignment of the principle axis (Dubinski 1992).  

At small separations, other factors, such as the recent history of
galaxy collisions, might also affect the ellipticity correlations.  In
addition, it is important to remember that we are probing only the
light distributions, which reflect the matter distribution only at the
very central parts of the galaxies.

It is essential to understand precisely how the ellipticity
correlations depend on the angular momenta.  The dominant contribution
to the correlations comes from alignments in the orientations of the
galaxy ellipticities.  The elongations of the galaxy light
distributions are expected to be orthogonal to the direction of their
projected angular momenta.  The magnitude of the ellipticity may to
some extent depend on the magnitude of $L$; for example, galaxies with
larger angular momenta may appear more disk-like.  Even so, the form
of this relation is largely irrelevant for understanding the
ellipticity correlations.  This is because the galaxy orientations are
expected to be isotropic on average, so there must be correlations in
the alignments for ellipticity correlations to occur.  However, the
magnitude of the ellipticity has a significant mean value. Therefore,
galaxy {\it alignments} already introduce ellipticity correlations
even in the absence of correlations in the {\it magnitude} of the
ellipticities.

To see this, consider the ellipticity correlation 
\begin{eqnarray} 
\langle \ep \ep'^{*} \rangle  \simeq \langle |\ep||\ep'| 
\rangle \langle e^{2i(\psi - \psi')} \rangle = 
[|\bar{\ep}|^2 + \langle (|\ep| - |\bar{\ep}|)(|\ep'| - |\bar{\ep}|) \rangle ] 
\langle e^{2i(\psi - \psi')}\rangle. 
\end{eqnarray} 
The first relation follows from assuming that the magnitudes of the
ellipticities are independent of their orientation correlations.
Recall that the distribution for ellipticities described in
Eqn. (\ref{eqn-p_e}) has a large mean value, $|\bar{\ep}| \simeq
0.42$.  The variance of this distribution is significantly smaller
than the square of the mean, $\langle (|\ep| - |\bar{\ep}|)^2 \rangle
= 0.055 \simeq 0.3 |\bar{\ep}|^2$, so that even perfect correlations
between magnitude of the ellipticities would only result in a small
modulation of the overall correlation.

Since the ellipticity is proportional to $e^{2i\psi}$, it is quadratic
in the angular momentum components perpendicular to the line of sight
(Eqn. (\ref{eqn-e_L}).)  The ellipticity correlation is therefore
quartic in the angular momenta: $\langle \ep \ep' \rangle \propto
\langle \hat{L}\hat{L}\hat{L}'\hat{L}' \rangle.$ This result should be
contrasted with the {\it Ansatz} of Catelan, Kamionkowski and
Blandford, which assumes the correlation to be quadratic in the
angular momenta.\footnote{These authors have recently re-examined this 
issue and now find results consistent with those we have presented here
[M. Kamionkowski, private communication.]}  

The correlation strength $\langle \epsilon\epsilon'\rangle$ also
depends on the mean ellipticity, which in turn depends on the galaxy
type.  Spiral galaxies are more flattened than elliptical galaxies,
and thus will have a larger correlation.  We have also assumed that
the angular momentum is parallel to the shortest axis of the galaxies,
which should be a good approximation for spirals, but may not be as
good for elliptical galaxies and could suppress their correlation
further.

Another major simplifying assumption we have made is that linear
theory is sufficient to calculate these angular momentum correlations.
We might hope that this is a good approximation, since most of the
angular momentum is expected to be imparted before the object starts
to collapse and enters the non-linear regime.  While there are
non-linear corrections, N-body simulations have shown the linear
approximation to be surprisingly robust (Lee and Pen 2000, Sugerman et
al. 2000).  We have attempted to account for the effects of non-linear
evolution by parameterizing the extent to which spin alignments are
suppressed in comparison with linear predictions.  This parameter was
estimated from the N-body simulations of Lee and Pen (2000) and
Heavens et al. (2000), as well as by comparison to measurements of
observed ellipticity correlations seen in the Tully catalog (Pen et
al. 2000).

The amplitude and shape of the ellipticity correlation function can be
understood intuitively. Recall that the ellipticity is a function of
the shear tensor, which is the second derivative of the potential.  By
virtue of Poisson's equation, the trace of the shear tensor is the
density.  Therefore we expect the correlation of the other components
of the shear field will drop at the same rate as the density
correlation function. Since the ellipticities are quadratic in the
shear field, correlations in them will fall as the density correlation
function {\it squared}, $\langle \epsilon\epsilon'\rangle\propto
\xi_{\rho}^2$.  The order-of-magnitude of the amplitude of the
correlation at zero-lag follows from simple symmetry arguments. The
shear tensor has six degrees of freedom, but only five are relevant
since angular momentum is independent of the trace. The typical
magnitude of a fourth order moment of a unit tensor in a five
dimensional space is 1/35. Therefore, these considerations suggest
that the ellipticity variance will have a comparable amplitude.
Fuller consideration shows that, from Equation(\ref{eqn:rough}),
$\langle \epsilon\epsilon'\rangle\propto 1/84\times
\xi^2(r)/\xi^2(r=0).$ Our analytic calculations are valid for random
points in a Gaussian field, but galaxies are usually assumed to form
at density peaks. We performed large realizations of Gaussian fields
and checked that our results are good approximations for such special
sampling.

We have compared the strength of the intrinsic correlation to that
expected for weak lensing. The intrinsic signal grows as the depth of
the survey decreases, because then galaxies close on the sky are on
average also physically closer together, hence they are more
correlated. The weak lensing signal, on the other hand, becomes
weaker, since there is less matter between us and the lensed
objects. For surveys typical of weak lensing, with a median redshift
of $z_m=1$, the intrinsic signal is of order of 1 per cent of the weak
lensing amplitude. However, for shallower surveys such as SDSS or 2dF,
the intrinsic signal may dominate the lensing one, on small
scales. Therefore, SDSS and 2dF are ideally suited for studying
intrinsic correlations in the orientations of galaxies.

The intrinsic ellipticity depends on the square of the tidal field,
whereas the lensing distortion is linear in the shear. As a direct
consequence, the distortion field is curl-free when induced by
lensing, but not when intrinsic correlations are present as well 
(Crittenden et al. 2000). 
The detection of such \lq magnetic\rq~ modes will be an
invaluable way of separating lensing from intrinsic correlations.

Finally, in this paper we have concentrated on intrinsic correlations
of galaxies.  Applying a similar reasoning to clusters, one could hope
to study the shear field on much larger scales.  The alignment of
clusters of galaxies is dominated by the intrinsic alignment of the
major shear axis.  Their dynamical time is longer, and they form
later, so we would expect the initial formation alignment to persist, 
implying ellipticities linearly proportional to the shear. 
The correlation should then drop as the correlation function $\xi(r)$
instead of its square as is the case for spin alignments.  The
qualitative features are reported for Gaussian random fields in Pen
(2000) and for simulations by Tseng and Pen (2000).


\begin{acknowledgments}
We thank L. van Waerbeke for useful conversations. 
RC and TT acknowledge PPARC for the award of an Advanced and a
post-doctoral fellowship, respectively. 
PN acknowledges support 
from a Trinity College Research Fellowship. 
Research conducted in
cooperation with Silicon Graphics/Cray Research utilizing the Origin
2000 supercomputer at the Department for Applied Mathematics and
Theoretical Physics (DAMTP), Cambridge. 
\end{acknowledgments}

\appendix 
\section{From intrinsic to projected shapes}
 
Here we derive an expression for the projected ellipticity,
$\epsilon\,=\,|\epsilon|\,e^{2i\psi}$, for a general ellipsoid 
when viewed from an arbitrary angle. We follow the treatment 
of Stark (1977) and consider a galaxy as an absorption-free 
stellar system, in which the volume brightness is constant 
on similar ellipsoids.

The general equation for the ellipsoid of the constant volume
brightness in the coordinate frame of the galaxy $(x,y,z)$ is,
\begin{eqnarray}
t^2\,x^2 +u^2\,y^2+z^2 = a_{v}^2, 
\end{eqnarray}
where $t$ is the axis ratio $c/a$, $u$ the axis ratio $c/b$ and
$a_{v}$ is the variable that parameterizes the volume brightness.
(Note that one axis ratio $u$ used here differs from the one used by
LBL, which is $b/a$.)  We want to transform the above equation into a
frame that is aligned with $z^{'}$ along the line of sight, which is
accomplished by a general rotation, characterized by the first two
Euler angles $\phi$ and $\theta$,
\begin{displaymath}
\left[\begin{array}{c} x\\y\\ z\\\end{array}\right]=\,\left[\begin{array}{ccc}
\cos {\phi}&{-\sin{\phi}\,\cos\theta}&{\sin{\phi}\,\sin\theta}\\
\sin {\phi}&{\cos{\phi}\,\cos{\theta}}&{-\cos{\phi}\,\sin{\theta}}\\
0 & \sin{\theta}&{\cos{\theta}}\\
\end{array}\right]\,\left[\begin{array}{c} x'\\y'\\z'\\
\end{array}\right].
\end{displaymath}

Stark shows that projecting the volume brightness along the line of
sight yields curves of constant surface brightness described by,
\begin{eqnarray}
a_s^2 = ({\frac{j}{f}}){x'^{2}} + 2({\frac{k}{f}}){x'}{y'} +
(\frac{l}{f}){y'^{2}}, 
\end{eqnarray}
where $a_s$ parameterizes the surface brightness and, 
\begin{eqnarray}
f &\equiv& f(\phi,\theta,t,u) \equiv t^2 \sin^2 \theta \sin^2 \phi + u^2
\sin^2 \theta \cos^2 \phi + \cos^2 \theta \\
j &\equiv& j(\phi,\theta,t,u) \equiv t^2 u^2 \sin^2 \theta + t^2 \cos^2
\phi \cos^2 \theta + u^2 \sin^2 \phi \cos^2 \theta \\
k &\equiv& k(\phi,\theta,t,u) \equiv (u^2 - t^2) \sin \phi \cos \phi
\cos \theta \\
l &\equiv& l(\phi,\theta,t,u) \equiv t^2 \sin^2 \phi + u^2 \cos^2 \phi. 
\end{eqnarray}
For a given set of axes ratios and observation angle, $j,k,l$ and $f$
are constant, so that the projection of curves with constant surface
brightness are similar ellipses. Therefore the projected image of a 
galaxy which has luminosity constant on similar ellipsoids has isophotes which
are similar ellipses, with the same position angle.

These isophotes correspond to ellipses with 
\begin{eqnarray} 
\beta^2 = \frac{1 - \sqrt{1-\gamma}}{1 + \sqrt{1-\gamma}} = (\frac{1 -
\epsilon}{1 + \epsilon})
\end{eqnarray}
where $\gamma = {4 t^2 u^2 f}/(j+l)^2$, and $\beta$ the ratio of the
short axis to the long axis, and
\begin{eqnarray} 
\psi = {1 \over 2} \sin^{-1}\left({\frac{2 k}{(j + l) \epsilon}}\right),
\end{eqnarray} 
the angle between the major axis and the $x'$
direction.  Therefore the general expression for the projected
ellipticity of a galaxy with axes ratios $t$ and $u$ seen from a
line of sight ($\theta, \phi$) with respect to the galaxy frame is,
$\ep(\theta, \phi, t, u) = \sqrt{1-\gamma} e^{2i\psi}.$

\section {Moments of $\hat{L}$}

As discussed in the text, the variance of the expectation value 
of the direction of the angular momentum is 
\begin{eqnarray}
\langle \hat{L}_\alpha \hat{L}_\beta \rangle = \int d^2\hat{L} \, \, 
\hat{L}_\alpha \hat{L}_\beta {1 \over 4\pi |{\cal{Q}}|^{1 \over 2}} 
(\hat{L}_\alpha {\cal{Q}}_{\alpha \beta}^{-1} \hat{L}_\beta)^
{-{3 \over 2}}. 
\end{eqnarray}  
Here, we estimate it in the limit of $ a \ll 1.$ A similar discussion
can be found in LP00. Writing the measure $d^2\hat{L} = d\hat{L}_1 d\phi$
the angular integral in the above expression can be evaluated
explicitly. In the frame where ${\cal{Q}}$ is diagonal, 
\begin{eqnarray}
\langle \hat{L}_1 \hat{L}_1 \rangle = \int_{-1}^{1}d\hat{L}_1
\int_{0}^{2 \pi} d\phi {1 \over 4\pi |{\cal{Q}}|^{1 \over 2}}\,
{\frac{\hat{L}_1^2}{(A \hat{L}_1^2 + B)^{3 \over 2}}},
\end{eqnarray}
where
\begin{eqnarray}
B &=& {\cal{Q}}_{22}^{-1} \cos^2 \phi + {\cal{Q}}_{33}^{-1} \sin^2 \phi
\nonumber \\ A &=& {\cal{Q}}_{11}^{-1} - B.
\end{eqnarray}
Substituting $\tan^2\psi = {A \over B} \hat{L}_1^2$, we have,
\begin{eqnarray}
\langle \hat{L}_1 \hat{L}_1 \rangle = 2
\int_{0}^{\tan^{-1}{\sqrt{A/B}}} d\psi \int_{0}^{2 \pi} d\phi 
{1 \over 4\pi |{\cal{Q}}|^{1 \over 2}}\,{\frac{\sin^2 \psi}{A^{3 \over
2} \cos \psi}}.
\end{eqnarray}
The $\psi$ integral can be evaluated exactly to be
\begin{eqnarray} 
\int d \psi {\frac{\sin^2 \psi}{\cos \psi}} = -\sin \psi + \log\left
[{\frac{1 + \tan{\psi/2}}{1 - \tan{\psi/2}}}\right] \sim 
\psi^3/3\,+\,\psi^5/30.
\end{eqnarray}
The approximation is valid in the small angle limit $\tan^{-1}
\sqrt{A/B} \ll 1$, which corresponds to assuming that $a$ in eqn. (9)
is small.  Using $\tan^{-1}\psi \simeq \psi - \psi^3/3$, the integral
becomes,
\begin{eqnarray} 
\langle \hat{L}_1 \hat{L}_1 \rangle \simeq {\frac{2}{3}} 
\int_{0}^{2 \pi} d\phi {1 \over 4\pi |{\cal{Q}}|^{1 \over 2}}\,
[\,B^{-{3 \over 2}}\,-\,{\frac{9}{10}}\,A\,{B^{-{5 \over 2}}}],
\end{eqnarray}
since, 
\begin{eqnarray} 
{\cal{Q}}_{\alpha \beta}^{-1} \simeq {{1-a} \over 3}\delta_{\alpha \beta} 
+ a \hat{T}_{\alpha \gamma} \hat{T}_{\gamma \beta},
\end{eqnarray}
we have,
\begin{eqnarray} 
A\,&=&\,a\,(\,\hat{T}_{11}^2\,-\,\hat{T}_{22}^2\,\cos^2\phi\,-\,\hat{T}_{33}\,\sin^2
\phi\,) \nonumber \\ B\,&=&\,{\frac{1 - a}{3}}\,+\,a \hat{T}_{22}^2
\cos^2 \phi + a \hat{T}_{33}^2 \sin^2 \phi.
\end{eqnarray}
To linear powers in $a$, the first term in the integral 
becomes,
\begin{eqnarray}
{\frac{2}{3}} \int_{0}^{2 \pi} d\phi {1 \over 4\pi |{\cal{Q}}|^{1
\over 2}}\, \,B^{-{3 \over 2}} &\simeq& {\frac{2}{3}} \int_{0}^{2 \pi}
d\phi {1 \over 4\pi}\,(1 - {\frac{3}{2}}(-a + 3 a [\hat{T}_{22}^2
\cos^2 \phi + \hat{T}_{33}^2 \sin^2 \phi])) \nonumber \\ &\simeq&
{\frac{1}{6 \pi}}[2\pi(1 + {\frac{3}{2}}a) - \frac{9 \pi
a}{2}(\hat{T}_{22}^2 + \hat{T}_{33}^2)] \nonumber \\ &\simeq&
{\frac{1}{3}}(1 - {\frac{3 a}{4}} + {\frac{9 a}{4}} \hat{T}_{11}^2).
\end{eqnarray}
Similarly, the second term in equation (B6) gives,
\begin{eqnarray}
{\frac{-3}{5}} \int_{0}^{2 \pi} d\phi {1 \over 4\pi |{\cal{Q}}|^ {1
\over 2}}\,\,A\,B^{-{5 \over 2}} &\simeq& {\frac{-3}{5}} \int_{0}^{2
\pi} d\phi {1 \over 4\pi } {1 \over 3}a\,(\,\hat{T}_{11}^2\,-\,
\hat{T}_{22}^2\,\cos^2\phi\,-\,\hat{T}_{33}^2\,\sin^2 \phi\,)
\nonumber \\ &\simeq& {\frac{-a}{20 \pi}}[2\pi \hat{T}_{11}^2 + \pi
( \hat{T}_{22}^2 + \hat{T}_{33}^2)] \nonumber \\ &\simeq& {a \over 15}
({3 \over 4} - {9 \over 4}\hat{T}_{11}^2\,).
\end{eqnarray}
Thus, $\langle \hat{L}_1 \hat{L}_1 \rangle \simeq {\frac{1}{3}}(1 -
{\frac{3 a}{5}} + {\frac{9 a}{5}} \hat{T}_{11}^2).$ Similar
expressions hold for the other diagonal correlations and the
off-diagonal elements remain zero.  Thus the full correlation matrix
becomes,
\begin{eqnarray}
\langle \hat{L}_{\alpha} \hat{L}_{\beta} \rangle = Q_{\alpha \beta} =
{\frac{1}{3}}(1 - {\frac{3 a}{5}}) \delta_{\alpha \beta} + {\frac{3
a}{5}} \hat{T}_{\alpha \gamma} \hat{T}_{\gamma \beta}.
\end{eqnarray}

\section{The linear and quadratic shear two-point functions} 

Here we perform integrations useful in evaluating the two and four
point functions of $\hat{\mathbf{T}}$.  As described in the text, we
will transform to a new basis, ${\mathbf{\cal{T}}} =
( (T_{11}+T_{22}+T_{33}) /\sqrt{3}, (T_{11} - T_{22})/\sqrt{2},
(T_{11} + T_{22} - 2T_{33})/\sqrt{6}, \sqrt{2}T_{12}, \sqrt{2}T_{13},
\sqrt{2}T_{23}).$ This is a convenient basis to integrate over the
trace (since it is irrelevant to the angular momentum) and because the
correlation function has a particularly simple form.  The
transformation matrix between these bases is:
\begin{eqnarray}
{\cal{T}} \equiv R T = \left[ \begin{array}{cccccc} 
\frac{1}{\sqrt{3}} &  \frac{1}{\sqrt{3}} & \frac{1}{\sqrt{3}} & 0 & 0& 0 \\
\frac{1}{\sqrt{2}} &  -\frac{1}{\sqrt{2}} & 0 & 0 & 0& 0 \\
\frac{1}{\sqrt{6}} &  \frac{1}{\sqrt{6}} & -\frac{2}{\sqrt{6}}& 0 & 0& 0 \\
0 & 0 & 0 & \sqrt{2} & 0 & 0 \\
0 & 0 & 0 & 0 & \sqrt{2} & 0 \\
0 & 0 & 0 & 0 & 0 & \sqrt{2} \\ \end{array} \right] 
\left[ \begin{array}{c} T_{11} \\ T_{22} \\ T_{33} \\ T_{12} \\ T_{13} \\T_{23} 
\end{array} \right]
\end{eqnarray}
We will work in this basis throughout this appendix.

Let us consider first the correlation function at zero separation in
this new basis.  In terms of its original indices, $C_0 = \frac{1}{15}
\xi(0) [\delta_{\alpha \beta}\delta_{\gamma \sigma}\,
+\,\delta_{\alpha \gamma} \delta_{\beta \sigma}\,+\,\delta_{\alpha
\sigma} \delta_{\beta \gamma}].$ In the new basis this becomes
\begin{eqnarray}
[{\cal{C}}_0]_{AB} =  R_{AA'}[C_0]_{A'B'} [R^{T}]_{B'B} = 
{\xi(0) \over 15} {\rm diag}(5, 2, 2, 2, 2, 2). 
\end{eqnarray} 
The factor in the exponential of the Gaussian distribution, $T
C^{-1}_0 T$, can be written in this basis as 
${\cal{T C}}^{-1}_0 {\cal {T}} = \xi^{-1}(0) ({\rm Tr}\,T^2 + 15
|{\cal{T}}|^2/2)$, where $|{\cal{T}}|^2 = \sum_{A=2}^6 {\cal{T}}^2_A$
is the modulus of the traceless part of ${\cal{T}}$.

In this basis, it is simple to calculate $\langle{\hat
{\cal{T}}_A}\,{\cal{T}}_B \rangle$ as
\begin{eqnarray}
\langle \hat{\cal{T}}_A\,{\cal{T}}_B \rangle
\,= \, \int {\frac{d^6 {\cal{T}}}{(2\pi)^3 
|{\cal{C}}_0|^{1/2}}}
{\hat {\cal{T}}_A}\,{\cal{T}}_B
\,e^{{-\frac{1}{2}[\xi^{-1}(0)({\rm Tr}\,T^2 + 15 |{\cal{T}}|^2/2)]}}.
\end{eqnarray}
Converting the measure to $d^6 {\cal{T}} = \frac{1}{\sqrt{3}} d {\rm
Tr}T |{\cal{T}}|^4 d |{\cal{T}}| d^4 {\hat{\cal{T}}}$ and rewriting
${\cal{T}}_D = |{\cal{T}}| {\hat{\cal{T}}}_D + {\rm Tr}\,T\,
\delta_{1D}/\sqrt{3}$ we can easily perform the integrations.  The
trace integral yields
\begin{eqnarray}
\int_{-\infty}^\infty d{\rm Tr}T 
e^{{-\frac{1}{2}[\xi^{-1}(0)({\rm Tr}\,T^2)]}}  = 
(2 \pi \xi(0))^{\frac{1}{2}}, 
\end{eqnarray}
while the modulus integral is
\begin{eqnarray}
\int_{0}^\infty |{\cal{T}}|^5 d |{\cal{T}}| 
e^{{-\frac{1}{2}[15 \xi^{-1}(0) |{\cal{T}}|^2/2)]}}  = 8 (2 \xi(0)/15)^3. 
\end{eqnarray}
The determinant in this basis is simply $|{\cal{C}}_0|^{1/2} =
4\sqrt{10} (\xi(0)/15)^3$ so that we find
\begin{eqnarray}
\langle \hat{\cal{T}}_A\,{\cal{T}}_B \rangle
&=& \, \frac{2}{\pi^{5/2}} \left({ \xi(0)\over 15}\right)^{1 \over 2}
\int d^4 {\hat{\cal{T}}} 
\, {\hat {\cal{T}}_A}\,{\hat{\cal{T}}}_B  = 
 \, \frac{2}{\pi^{5/2}} \left({ \xi(0)\over {15}}\right)^{1 \over 2}
 \frac{8 \pi^2}{15} \delta_{AB}\,=\,\frac{16}{15 \pi^{1/2}} 
\left({\xi(0)\over {15}}\right)^{1 \over 2}
\delta_{AB},   
\end{eqnarray}
where $\delta_{AB}$ runs only over the non-trace indices (2-6).  
Effectively, operating $\langle \hat{\cal{T}}_A\,{\cal{T}}_B \rangle$ 
on a vector projects out the trace part of the vector.  
We are finally in a position to find the linear two point function:
\begin{eqnarray}
\langle {\hat {\cal{T}}_{A}\,{\hat {\cal{T}}_{B}}}'
\rangle\,&=& \langle {\hat{\cal{T}}_A}\,{{\cal{T}}_C}
\rangle\,[{\cal{C}}^{-1}_0 {\cal{C}}_{r} {\cal{C}}^{-1}_0]_{CD}\,\langle 
{\hat {\cal{T}}_B'}\,{{\cal{T}}_D}' \rangle \nonumber \\ & = & \frac{64}{15\pi
\xi^(0)} \delta_{AC} [{\cal{C}}_{r}]_{CD} \delta_{DB} =
\frac{64}{15\pi\xi(0)} [\tilde{\cal{C}}_{r}]_{AB},
\end{eqnarray} 
where the tilde denotes that the trace has been projected out of the
correlation function. (If $P$ is the projection operator, then
$\tilde{C} \equiv P C P^T$.  In the original basis, this projection
operator is $R^{-1} \delta_{AB} R$.)  
At zero
separation, this gives an answer to within 10\% of the exact
value $\langle {\hat {\cal{T}}_{A}\,{\hat {\cal{T}}_{B}}}' \rangle\, =
\delta_{AB}/5$, a remarkable fact when one remembers that this was
derived assuming $C_r \ll C_0$.

Moving on, we next try to evaluate the quadratic two-point correlation
in this basis,
\begin{eqnarray}
\langle {\hat {\cal{T}}_{A}}\,{\hat {\cal{T}}_{B}} {\hat {\cal{T}}_{C}}' {\hat {\cal{T}}_{D}}'
\rangle\,&=& \,\int\,\frac{d^6 {\cal{T}} d^6 {\cal{T}}'}
{{|{\cal{C}}|^{1/2}}{(2\,\pi)^6}}
{\hat {\cal{T}}}_{A}\,{\hat {\cal{T}}_{B}}{\hat {\cal{T}}_{C}}' {\hat {\cal{T}}_{D}}' \,e^
{-\frac{1}{2}{\vec {\cal{T}}^{{\cal{T}}}}{{\cal{C}}^{-1}}{\vec {\cal{T}}}}\, \nonumber \\ 
&=& \langle {\hat {\cal{T}}_{A}}\,{\hat {\cal{T}}_{B}}\rangle \langle 
{\hat {\cal{T}}_{C}}' {\hat {\cal{T}}_{D}}'
\rangle\,+\frac{1}{2} \langle {\hat {\cal{T}}_{A}}\,{\hat {\cal{T}}_{B}} {{\cal{T}}_{E}}
{{\cal{T}}_{F}} \rangle\, \langle {\hat {\cal{T}}_{C}}\,{\hat {\cal{T}}_{D}} {{\cal{T}}_{G}}
{{\cal{T}}_{H}} \rangle\,[{\cal{C}}^{-1}_0\,{\cal{C}}_{\mathbf r}\,{\cal{C}}^{-1}_0]_{EG}
[{\cal{C}}^{-1}_0\,{\cal{C}}_{\mathbf r}\,{\cal{C}}^{-1}_0]_{FH}.
\end{eqnarray}
Now, at zero-lag the quartic moment in the transformed basis,
\begin{eqnarray}
\langle {\hat {\cal{T}}_A}\,{\hat {\cal{T}}_B} {\cal T}_{C}{\cal T}_{D}
\rangle\,= \, \int {\frac{d^6 {\cal{T}}}{(2\pi)^3 
|{\cal{C}}_0|^{1/2}}}
{\hat {\cal T}_{A}}\,{\hat {\cal{T}}_{B}} {\cal T}_{C}{\cal T}_{D}
\,e^{{-\frac{1}{2}[\xi^{-1}(0)({\rm Tr}\,T^2 + 15 |{\cal{T}}|^2/2)]}}.
\end{eqnarray}
Again making the substitution ${\cal T}_{C} = {\cal T}{\hat {\cal
T}_{C}} + \frac{1}{\sqrt{3}} {\rm Tr}\,T\,\delta_{1C}$, the surviving 
terms are of the following form,
\begin{eqnarray}
\langle {\hat {\cal T}_A}\,{\hat {\cal T}_B} {\cal T}_{C}{\cal T}_{D}
\rangle\,= \int \frac{\frac{1}{\sqrt{3}}
d {\rm Tr}T |{\cal{T}}|^4 d |{\cal{T}}| d^4 {\hat{\cal{T}}}}{(2\pi)^3 
|{\cal{C}}_0|^{1/2}}{\hat {\cal T}_{A}}\,{\hat {\cal{T}}_{B}}[|{\cal
T}|^2 {\hat {\cal T}_{C}} {\hat {\cal T}_{D}} + \frac{{\rm Tr T}^2}{3}
\delta_{1C} \delta_{1D}].
\end{eqnarray}

As above we can perform the integrals simply.  For the first term, the
trace integral is identical to equation (C5), while the
modulus integral is
\begin{eqnarray}
\int_{0}^\infty |{\cal{T}}|^6 d |{\cal{T}}| 
e^{{-\frac{1}{2}[15 \xi^{-1}(0) |{\cal{T}}|^2/2]}}  = 15
\frac{\sqrt{2 \pi}}{2}(2 \xi(0)/15)^{\frac{7}{2}}. 
\end{eqnarray} 
The angular integral yields  
\begin{eqnarray}
\int d^4{\hat {\cal T}} {\hat {\cal T}_{A}} {\hat {\cal T}_{B}} {\hat
{\cal T}_{C}} {\hat {\cal T}_{D}} = \frac{8 \pi^2}{105}[\delta_{AB}
\delta_{CD} + \delta_{AC} \delta_{BD} + \delta_{AD} \delta_{BC}].
\end{eqnarray}
Again the indices range over 2-6, since the trace has been 
effectively projected out. 
For the second term, the angular integral is as it was for the linear 
two point function (C6), while the trace integral becomes
\begin{eqnarray}
\int_{-\infty}^\infty d\,{{\rm Tr} T}\,({{\rm Tr}\,T})^2 
e^{{-\frac{1}{2}[\xi^{-1}(0)({\rm Tr}\,T^2)]}}  
= \sqrt{2 \pi} \xi(0)^{\frac{3}{2}}, 
\end{eqnarray}
and the modulus integral gives,
\begin{eqnarray}
\int_{0}^\infty |{\cal{T}}|^4 d |{\cal{T}}| 
e^{{-\frac{1}{2}[15 \xi^{-1}(0) |{\cal{T}}|^2/2]}}  = \frac{3 \sqrt{2
\pi}}{2}(2 \xi(0)/15)^{\frac{5}{2}}. 
\end{eqnarray}
Putting these all together, we find that 
\begin{eqnarray}
\langle {\hat {\cal{T}}_A}\,{\hat {\cal{T}}_B} {\cal T}_{C}{\cal T}_{D}
\rangle\,= {\xi(0)\over 15} \left [\frac{2}{7}(\delta_{AB}  
\delta_{CD} + \delta_{AC} \delta_{BD} + \delta_{AD}  \delta_{BC}) 
+ \delta_{AB} \delta_{1C} \delta_{1D}\right]. 
\end{eqnarray}

Finally, we are in a position to evaluate the quadratic two point function.
This expression has a number of terms, but it can effectively be
broken into a local part, which includes terms
proportional to $\delta_{AB}$ or $\delta_{CD}$, and a non-local part.
Since only the latter terms contribute to the ellipticity correlation,
we will keep only these here. This non-local part is a simple
function of the correlation of the trace-free components,  
\begin{eqnarray}
\langle {\hat {\cal{T}}_{A}}\,{\hat {\cal{T}}_{B}} {\hat
{\cal{T}}_{C}}' {\hat {\cal{T}}_{D}}' \rangle\, &=& \left(\frac{15}{14
\xi(0)} \right)^2 \left[ \tilde{\cal{C}}_{AC} \tilde{\cal{C}}_{BD} +  
\tilde{\cal{C}}_{AD}  \tilde{\cal{C}}_{BC} \right]\,+\,{\rm local\, terms}.
\end{eqnarray}
In the limit of small separations, the total (local and non-local)
correlation function should approach $[\delta_{AB}
\delta_{CD} + \delta_{AC} \delta_{BD} + \delta_{AD} \delta_{BC}]/35$.
\end{document}